\documentclass[preprint,10pt]{elsarticle}
\usepackage{amssymb}
\usepackage{lmodern}
\usepackage{amsmath}
\usepackage[utf8]{inputenc} 
\usepackage[T1]{fontenc}    
\usepackage{hyperref}       
\usepackage{url}            
\usepackage{booktabs}       
\usepackage{amsfonts}       
\usepackage{nicefrac}       
\usepackage{microtype}      
\usepackage{lipsum}
\usepackage{afterpage}
\usepackage{graphicx,xcolor}
\usepackage{float}
\usepackage[ruled]{algorithm2e}
\usepackage[cal=boondox]{mathalfa}
\usepackage[mathscr]{eucal}
\usepackage[ruled]{algorithm2e}
\usepackage{breqn}
\usepackage{array}
\usepackage{makecell}
\usepackage{multicol}
\usepackage{setspace}

\begin{document}
\begin{frontmatter}
\title{Emergence of group hierarchy}

\author[LISC,LAPSCO]{Guillaume Deffuant}
\ead{guillaume.deffuant@inrae.fr}
\author[LISC]{Thibaut Roubin}

\address[LISC]{Université Clermont Auvergne, INRAE, UR LISC, Aubi\`ere, France}
\address[LAPSCO]{Université Clermont Auvergne, LAPSCO, Clermont-Ferrand, France.}

\maketitle
\begin{abstract}
We consider an opinion dynamics model where, during random pair interactions, each agent modifies her opinions about both agents of the random pair and also about some other agents, chosen randomly. Moreover, each agent belongs to a single group and the opinions within the group are attracted to their average. In simulations starting from neutral opinions, we observe the emergence of a group hierarchy. We derive a moment approximation that provides equations ruling the evolution of the average opinion of agents in a group about the agents of another group. This approximation explains how the group hierarchy emerges.
\end{abstract}


\end{frontmatter}

\section{Introduction}

This paper builds on previous work \cite{Deffuant2013,Deffuant2018,Deffuant2022} on an opinion dynamics model in which opinions are about the agents themselves instead of being about products or political options. 

Indeed in most opinion dynamics models  (\cite{French1956,Galam2002,Deffuant2000,Deffuant2006a,Hegselmann2002,Flache2011}, for a recent review see: \cite{Flache2017}), apart from a few exceptions \cite{Bagnoli2007,Carletti2011,Flache2018}, opinions about the agents themselves are not considered as deserving any specific attention. However, the opinions about agents determine the social network of positive or negative connections, hence in some respect the social structure.  Moreover, it is generally recognised that this social structure has a strong influence on the agents' opinions. This suggests that opinions about agents do matter. Several opinion dynamics models include such a structure and in some cases it is evolving. This is for instance the case of some versions of the social impact model \cite{Latane1981,Holyst2001}. Moreover, other researches propose models of social structure dynamics, for instance hierarchies resulting from fights between primates \cite{Bonabeau1996}. In both cases, the social structure is generated by processes that are different from opinion dynamics.

In this paper, we assume that the agents of the model proposed in \cite{Deffuant2018,Deffuant2022} belong to different groups. The type of groups that we have in mind would be male and female or other physical differences that can easily be observed (e.g. colour of skin).
We keep all the assumptions made in the previous work. The agents hold an opinion (a real number between -1 and +1) about all the others and themselves. The model dynamics repeats encounters of two randomly chosen agents influencing their self-opinions and their opinions about each other. Moreover, if gossip is activated, both agents influence their opinions about some other randomly chosen agents. In all cases, the influence is attractive and the agents are more strongly influenced by the ones that they hold in high esteem. Importantly, agents do not have a transparent access to the opinions of others; they constantly make errors of interpretation which are modelled by a random noise.

The only novelty in the dynamics is the introduction of some group conformity: the opinion of an agent of group $I$ about an agent of group $J$ is slightly attracted to the average opinion of agents of group $I$ about agents of group $J$. The average opinion in $I$ about group $J$ hence plays the role of a prejudice of agents of group $I$ about the agents of group $J$. 

In the same line as in \cite{Deffuant2022}, we develop an analytical approximation of the average (first moment) evolution of the opinions in the model and the average evolution of their products (second moment), following a general approach \cite{Law2000}. The approximation goes one step further because it considers the average opinion of the agents of group $I$ about the agents of group $J$. 

The analysis of the approximate model shows the existence of biases on average group opinions that are similar to the biases on opinions about agents that are analysed in \cite{Deffuant2022}. The structure of these biases explains that the opinions about the groups of low status tend to decrease while the opinions about groups of high status tend to increase. 

The following section describes the model and presents the patterns in more details. Section 3 is devoted to the moment approximation. Section 4 analyses the accuracy of the approximation and studies the effect of initial group hierarchies. The last section discusses the results and their limitations. 

\section{The model}
\subsection{State}

The model includes $n_g$ groups of $N_g$ agents. Each agent $i \in \{1,\dots, n_g N_g\}$ belongs to a single group $I \in \{1,\dots, n_g\}$ and has an opinion $a_{ij}$ about each agent $j \in \{1,\dots, N_g\}$  including herself. The opinions are real values between -1, the worst opinion, and +1, the best opinion.  
\subsection{Dynamics}

At each time steps, two randomly chosen distinct agents  $i$ and $j$ encounter and discuss their opinions about each other. These discussions generate only a part of all the opinion changes occurring in a time step and we assume that these changes take place at $t + 0.5$ (the next changes are the attraction to the average group opinion and we assume that they take place at $t +1$). We have:

\begin{align}
    a_{ii}(t+0.5) &= a_{ii}(t) + h_{ij}(t)(a_{ji}(t) - a_{ii}(t) + \theta_{ii}(t)),\\
    a_{ji}(t+0.5) &= a_{ji}(t) + h_{ji}(t)(a_{ii}(t) - a_{ji}(t) + \theta_{ji}(t)),
\end{align}

where $\theta_{ij}(t)$ is a random number uniformly drawn in $[-\delta, \delta]$ which represents errors in the evaluation of others' opinions. The function $h_{ij}(t)$ represents the influence that agent $j$ has on agent $i$. This influence is high if $i$ has a high opinion of $j$ and low otherwise. The expression of  $h_{ij}(t)$ is: 

\begin{equation}\label{eq:credibility}
    h_{ij}(t) = H(a_{ii}(t) - a_{ij}(t)) = \frac{1}{1 + \exp{\left( \frac{ a_{ii}(t) - a_{ij}(t) }{\sigma}\right)} }, 
\end{equation}

where $\sigma$ is a parameter of the model. The same modifications are made on the opinions $a_{jj}$ and $a_{ij}$.

Then, if gossip is activated, the two agents $i$ and $j$ discuss their opinions about $k > 0$ randomly chosen other agents $g$, distinct from $i$ and from $j$:
\begin{align}
    a_{ig}(t+0.5) &= a_{ig}(t) + h_{ij}(t)(a_{jg}(t) - a_{ig}(t) + \theta_{ig}(t)).
\end{align}

After the interaction occurring at $t+0.5$, each opinion is attracted by the average opinion of the agents of the same group at $t+1$:
\begin{align}
    a_{ii}(t+1) &= \mu.a_{ii}(t+0.5) + \frac{1-\mu}{N_g}\sum_{p\in I}a_{pp}(t+0.5),\\
    a_{ji}(t+1) &= \mu.a_{ji}(t+0.5) + \frac{1-\mu}{N_g(N_g - \delta_{IJ})}\sum_{p\in I}\sum_{q\in J,q \neq p} a_{qp}(t+0.5),
\end{align}

where $N_g$ the number of agents in each group.
\subsection{Observed patterns}
\label{sec:patterns}
\begin{figure}[!htb]
\centering
    \begin{tabular}{cc}
	  \includegraphics[width= 6 cm]{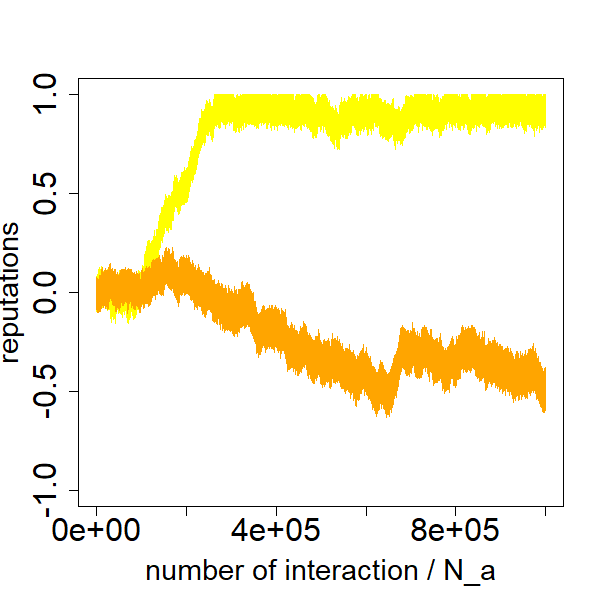}  & \includegraphics[width= 6 cm]{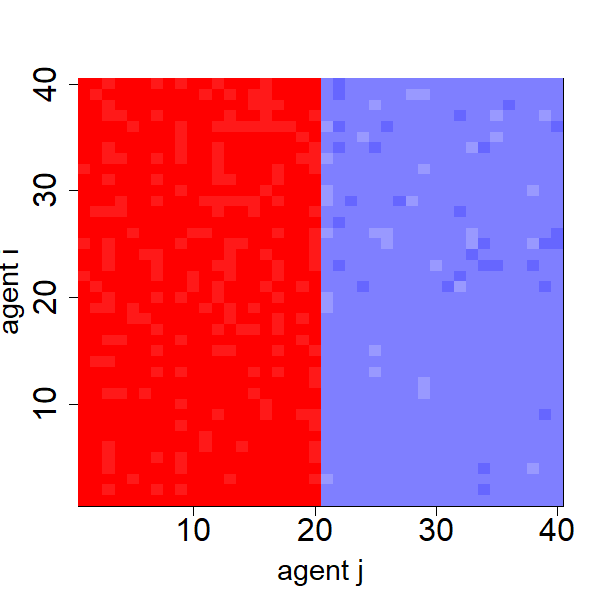} \\[-5pt]
	  \includegraphics[width= 6 cm]{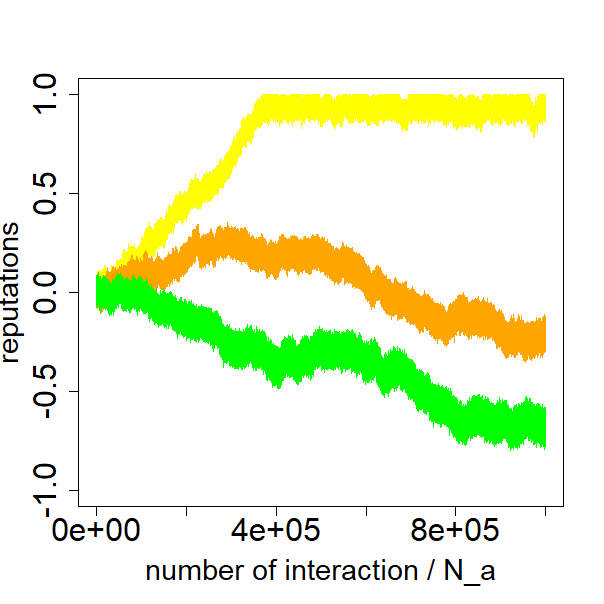}  & \includegraphics[width= 6 cm]{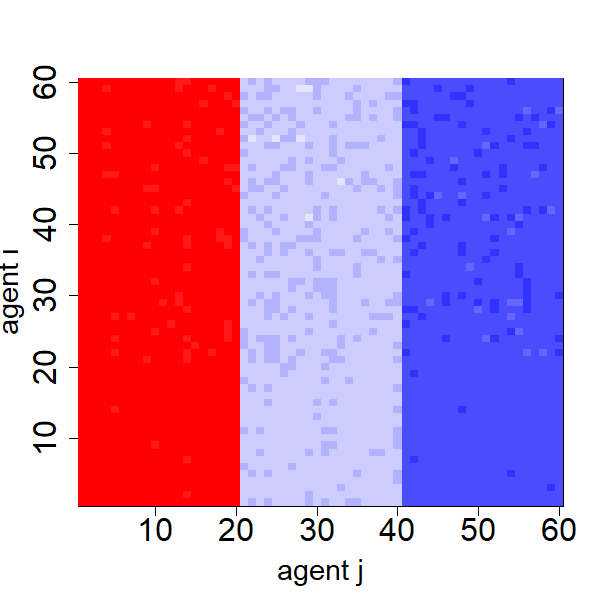}\\[-5pt]
	\end{tabular}
   \caption{Example opinion evolution (left) and opinion matrix after $N_a$ million pair interactions (right), for 2 groups (top) and 3 groups (bottom). In the matrix, blue and red squares represent respectively negative and positive opinions. The lighter the colour, the closer to 0 (neutral opinion) is the represented opinion. Agents 1 to 20 are in group 1, 21 to 40 in group 2, and for the case with 3 groups, agents 41 to 60 are in group 3. Noise $\delta = 0.05$, influence parameter $\sigma = 0.3$, attraction to group average $\mu = 0.995$.}
	\label{fig:opMatrix}
\end{figure}

Starting from all opinions equal 0, when gossip is activated, we observe the emergence of a rather stable group hierarchy. Figure  \ref{fig:opMatrix} shows a typical example of opinion matrices after a large number of interactions and of the evolution of the opinions over time. Rapidly a group hierarchy emerges. Note that all the agents of a group have similar opinions about the agents of any group even with a value of the parameter ruling the attraction of the opinions to the group average very close to $1$. This hierarchy is rather stable over time.

We propose now a mathematical analysis, based on the moment approximation technique, that explains these patterns.

\section{Moment approximation}

\subsection{Dynamics of the average opinion in a group about agents of a group}
\subsubsection{Equations of average group opinions.}
We set $x_{II}$ as the average self-opinion of agents of the group $I$ and $x_{JI}$ as the average opinion of agents of group $J$ on agents of group $I$. If $I = J$, we adopt the following notation: $x_{JI} = x_{II^j}$. This allows us to distinguish $x_{II}$, the average self-opinion of the agents of group $I$, from $x_{II^j}$, the average opinion of agents of group $I$ about other agents of group $I$. 

\begin{align} \label{eq:defxSI}
    x_{II}(t) = \frac{1}{N_g} \sum_{i \in I} x_{ii}(t).
\end{align}

\begin{align}  \label{eq:defxJI}
    x_{JI}(t) = \frac{1}{N_g (N_g - \delta_{IJ})} \sum_{(i,j) \in I \times J, i \neq j} x_{ji}(t).
\end{align}

For a given sequence of interactions $s_t$, assuming that the interacting couple at $t$ is $(i,j)$, let $x_{ii}(s_{t+0.5})$ be the value of $x_{ii}$ after the interaction and before the attraction towards the average :
\begin{align}
    x_{ii}(s_{t+0.5}) &= x_{ii}(s_t) + h_{ij}(s_t)(x_{ji}(s_t) - x_{ii}(s_t) + \theta_{ii}(s_t)).
\end{align}

Let $h_{IJ}(s_t) = H(x_{II}(s_t) + a_{II}(0) - x_{IJ}(s_t) - a_{IJ}(0)) $, $\overline{h_{IJ}}(t) = H(\overline{x_{II}}(t) + a_{II}(0) - \overline{x_{IJ}}(t) - a_{IJ}(0)) $ and $z_{IJ}(t) = x_{II}(t) - x_{IJ}(t)$. Developing $h_{ij}(s_t)$ around $\overline{h_{IJ}(t)}$ at the first order, we have:
\begin{align}
    h_{ij}(s_t) &= \overline{h_{IJ}(t)} + \overline{h'_{IJ}(t)}(x_{ii}(s_t) - x_{ij}(s_t) - \overline{z_{IJ}}(t)).
\end{align}
We simplify this equation to:
\begin{align}
    h_{ij}(s_t) &= \widehat{h_{IJ}}(t) + \overline{h'_{IJ}(t)}(x_{ii}(s_t) - x_{ij}(s_t)), 
\end{align}
using the notation:
\begin{align}
    \widehat{h_{IJ}}(t) = \overline{h_{IJ}(t)} - \overline{h'_{IJ}(t)} \overline{z_{IJ}}(t).
\end{align}
Replacing $h_{ij}(s_t)$ by its approximate value, we get:
\begin{dmath}
    x_{ii}(s_{t+0.5})  =  x_{ii}(s_t) + \widehat{h_{IJ}}(t)(x_{ji}(s_t) - x_{ii}(s_t) + \theta_{ii}(s_t))  + \overline{h'_{IJ}}(t) (x_{ii}(s_t) - x_{ij}(s_t))(x_{ji}(s_t) - x_{ii}(s_t) + \theta_{ii}(s_t)).
\end{dmath}

Then, averaging over all possible noise draws during the sequence $s_t$, we get:

\begin{dmath}
    \overline{x_{ii}}(s_{t+0.5})  =  \overline{x_{ii}}(s_t) + \widehat{h_{IJ}}(t)(\overline{x_{ji}}(s_t) - \overline{x_{ii}}(s_t))  +  \overline{h'_{IJ}}(t) (\overline{x_{ji}(s_t).x_{ii}(s_t)} - \overline{x^2_{ii}}(s_t) + \overline{x_{ij}(s_t).x_{ii}(s_t)}) - \overline{x_{ij}(s_t).x_{ji}(s_t)}.
\end{dmath}

Summing up over all agents in group $I$ and dividing by $N_g$, the number of agents in a group, we get:

\begin{dmath}
   \overline{x_{II}}(s_{t+0.5})  =   \overline{x_{II}}(s_t) + \frac{\widehat{h_{IJ}}(t)}{N_g}( \overline{x_{ji}}(s_t) -  \overline{x_{ii}}(s_t))  + \frac{\overline{h'_{IJ}}(t)}{N_g} (\overline{x_{ji}(s_t).x_{ii}(s_t)} - \overline{x^2_{ii}}(s_t) + \overline{x_{ij}(s_t).x_{ii}(s_t)}) - \overline{x_{ij}(s_t).x_{ji}(s_t)}.
\end{dmath}

Then, averaging over all sequences $s_t$, i.e. over all possibilities of couples $(i,j)$ each having probability $\frac{2}{N_c}$ to be drawn with $N_c = N_a(N_a-1)$, we get:

\begin{dmath}
    \overline{x_{II}}(t+0.5)  =  \overline{x_{II}}(t) + \sum_{J}\frac{2}{N_g N_c}\sum_{(i,j)\in I\times J,i \neq j} \left(\widehat{h_{IJ}}(t)(\overline{x_{ji}}(t) - \overline{x_{ii}}(t))  +  \overline{h'_{IJ}}(t) (\overline{x_{ji}(t).x_{ii}(t)} - \overline{x^2_{ii}}(t) + \overline{x_{ij}(t).x_{ii}(t)}) - \overline{x_{ij}(t).x_{ji}(t)})\right).
\end{dmath}

Similarly:
\begin{dmath}
    \overline{x_{ji}}(s_{t+0.5})  =  \overline{x_{ji}}(s_t) + \widehat{h_{JI}}(t)(\overline{x_{ii}}(s_t) - \overline{x_{ji}}(s_t))  + \overline{h'_{JI}}(t) (\overline{x^2_{ji}}(s_t) - \overline{x_{ii}(s_t).x_{ji}(s_t)} + \overline{x_{ii}(s_t).x_{jj}(s_t)} - \overline{x_{jj}(s_t).x_{ji}(s_t)}).
\end{dmath}

Hence, summing up over all sequences such that $i \in I$ and $j \in J$, and on $J \in G$ the possible groups, and dividing by $N_g(N_g - \delta_{IJ})$ (with $\delta_ {IJ} = 1$ \textit{iff} $I=J$) we get:

\begin{dmath}
    \overline{x_{JI}}(s_{t+0.5})  =  \overline{x_{JI}}(s_t) + \frac{1}{N_g(N_g - \delta_{IJ})}\left(\widehat{h_{JI}}(t)(\overline{x_{ii}}(s_t) - \overline{x_{ji}}(s_t))  + \overline{h'_{JI}}(t) (\overline{x^2_{ji}}(s_t) - \overline{x_{ii}(s_t).x_{ji}(s_t)}  + \overline{x_{ii}(s_t).x_{jj}(s_t)} - \overline{x_{jj}(s_t).x_{ji}(s_t)})\right).
\end{dmath}

Averaging over all the possible sequences $s_t$ hence all possible choices of $(i,j)$:
\begin{dmath}
    \overline{x_{JI}}(t+0.5)  =  \overline{x_{JI}}(t) + \frac{2}{N_c N_g(N_g - \delta_{IJ})}\sum_{(i, j) \in I \times J, i \neq j}\left(\widehat{h_{JI}}(t)(\overline{x_{ii}}(t) - \overline{x_{ji}}(t))  + \overline{h'_{JI}}(t) (\overline{x^2_{ji}}(t) - \overline{x_{ii}(t).x_{ji}(t)}  + \overline{x_{ii}(s_t).x_{jj}(s_t)} - \overline{x_{jj}(s_t).x_{ji}(s_t)})\right). 
\end{dmath}

We now introduce the average products of opinions at group level (second moments):
\begin{align}
    x^2_{IIII}(t) = \frac{1}{N_g}\sum_{i \in I } x_{ii}(t)x_{ii}(t);
\end{align}
\begin{align}
    x^2_{JIJI}(t) = \frac{1}{N_g(N_g - \delta_{IJ})} \sum_{(i,j) \in I \times J, i \neq j} x_{ji}(t)x_{ji}(t);
\end{align}
The other products $x^2_{IIJI}(t)$, $x^2_{IIJJ}(t)$, $x^2_{JIIJ}$, are defined similarly.

Then:
\begin{dmath}
    \overline{x_{II}}(t+0.5)  =  \overline{x_{II}}(t) + \sum_{J \in G}\frac{2(N_g - \delta_{IJ})}{N_c} \left(\widehat{h_{IJ}}(t)(\overline{x_{JI}}(t) - \overline{x_{II}}(t))  +  \overline{h'_{IJ}}(t) (\overline{x^2_{IIJI}}(t) - \overline{x^2_{IIII}}(t) + \overline{x^2_{IIIJ}}(t) - \overline{x^2_{IJJI}}(t))\right).
\end{dmath}

and:

\begin{dmath}
 \overline{x_{JI}}(t+0.5)  =  \overline{x_{JI}}(t) + \frac{2}{N_c}\left(\widehat{h_{JI}}(t)(\overline{x_{II}}(t) - \overline{x_{JI}}(t))  + \overline{h'_{JI}}(t) (\overline{x^2_{JIJI}}(t) - \overline{x^2_{IIJI}}(t) + \overline{x^2_{IIJJ}}(t) - \overline{x^2_{JJJI}}(t))\right).
\end{dmath}

Gossip modifies the equation of $\overline{x_{JI}}(t+0.5)$ as follows:

\begin{dmath}
    \overline{x_{JI}}(t+0.5)  =  \overline{x_{JI}}(t) + \frac{2}{N_c}\left(\widehat{h_{JI}}(t)(\overline{x_{II}}(t) - \overline{x_{JI}}(t))  + \overline{h'_{JI}}(t) (\overline{x^2_{JIJI}}(t) - \overline{x^2_{IIJI}}(t))+ \overline{x^2_{IIJJ}}(t) - \overline{x^2_{JJJI}}(t)\right) + \frac{2k}{N_t}\sum_{P \in G} (N_g - \delta_{IP} - \delta_{JP}) \widehat{h_{JP}}(t)(\overline{x_{PI}}(t) - \overline{x_{JI}}(t)),
\end{dmath}

with:

\begin{dmath}
    N_t  =  N_a(N_a-1)(N_a-2)
\end{dmath}

\subsubsection{Equations of average products}
In order to compute the evolution of the average group opinions, we need to compute evolution of average products of group opinions that appear in the equations and also the different products that may appear in the expressions of the products themselves. We show below the example of $ \overline{x^2_{IIII}}(t+0.5)$. 

\begin{dmath}
    \overline{x^2_{ii}}(t+0.5)  =  \overline{x^2_{ii}}(t) + \frac{2}{N_c} \sum_{J \in G} \sum_{j \in J, j \neq i} \left((\widehat{h^2_{IJ}}(t)-2\widehat{h_{IJ}}(t))\overline{x^2_{ii}}(t) + \widehat{h_{IJ}}(t)^2\overline{x^2_{ji}}(t) + 2 (1-\widehat{h_{IJ}}(t))\widehat{h_{IJ}}(t)\overline{x_{ii}(t)x_{ji}(t)} + \widehat{h_{IJ}}(t)^2 \frac{\delta^2}{3} \right).
\end{dmath}

Therefore:
\begin{dmath}
    \overline{x^2_{IIII}}(t+0.5)  = \overline{x^2_{IIII}}(t) + \sum_{J \in G}\frac{2(N_g - \delta_{IJ})}{N_c} \left((\widehat{h^2_{IJ}}(t)-2\widehat{h_{IJ}}(t))\overline{x^2_{IIII}}(t) + \widehat{h_{IJ}}(t)^2\overline{x^2_{JIJI}}(t) + 2 (1-\widehat{h_{IJ}}(t))\widehat{h_{IJ}}(t)\overline{x^2_{IIJI}(t)} + \widehat{h_{IJ}}(t)^2 \frac{\delta^2}{3} \right).
\end{dmath}
 Note that the expression $\frac{\delta^2}{3}$ appearing in the previous equation comes from the average of the squared noise $\overline{\theta^2}(t) = \frac{\delta^2}{3}$ (see \cite{Deffuant2022} for details).

A dozen of such products for which the similar equations are needed to complete the approximate model (it should be noted that more products should be considered than in the model without groups). We do not report them here for sake of paper concision. The interested reader can find the complete set of equations in the appendix.

\subsubsection{Attraction to the average.} After each interaction, the opinion modified by the interaction gets a bit closer to the average opinion:
\begin{align}
\label{eq:groupOp}
    \overline{x_{ii}}(t+1) = \mu.\overline{x_{ii}}(t+0.5) + (1-\mu).\overline{x_{II}}(t+0.5), \forall i\in \{1,\dots, N_a\}.
\end{align}

That does not change change the value of $x_{II}$, so finally, we obtain:
\begin{dmath}
    \overline{x_{II}}(t+1)  =   \overline{x_{II}}(t+0.5).
\end{dmath}

Similarly:
\begin{dmath}
    \overline{x_{JI}}(t+1)  =   \overline{x_{JI}}(t+0.5).
\end{dmath}

 However, in general, the attraction to the average modifies the expression of the products. For example, the equation of $\overline{x^2_{IIII}}(t+1)$ is computed as follows.

\begin{dmath}
   \overline{x_{ii}^2}(t+1)  =  \left(\mu\overline{x_{ii}}(t+0.5)+ \frac{1 - \mu}{N_g}\sum_{j \in I}\overline{x_{jj}}(t+0.5)\right)^2
\end{dmath}

\begin{dmath}
   \overline{x_{ii}^2}(t+1)  =  \mu^2\overline{x_{ii}^2}(t+0.5)+ \frac{2\mu(1 - \mu)}{N_g}\overline{x_{ii}}(t+0.5)\sum_{j \in I}\overline{x_{jj}}(t+0.5)
   + \frac{(1 - \mu)^2}{N_g^2}\sum_{j \in I}\sum_{p \in I}\overline{x_{jj}(t+0.5)x_{pp}(t+0.5)}
\end{dmath}

\begin{dmath}
   \frac{1}{N_g}\sum_{i \in I} \overline{x_{ii}^2}(t+1)  =  \frac{\mu^2}{N_g}\sum_{i \in I}\overline{x_{ii}^2}(t+0.5)+ \frac{2\mu(1 - \mu)}{N_g^2}\sum_{i \in I}\sum_{j \in I}\overline{x_{ii}(t+0.5)x_{jj}(t+0.5)}
   + \frac{(1 - \mu)^2}{N_g^2}\sum_{j \in I}\sum_{p \in I}\overline{x_{jj}(t+0.5)x_{pp}(t+0.5)}
\end{dmath}

Therefore:

\begin{dmath}
   \overline{x^2_{IIII}}(t+1)  = \left(\mu^2 + \frac{1-\mu^2}{N_g} \right)  \overline{x^2_{IIII}}(t+0.5)+ \frac{(1-\mu^2)(N_g-1)}{N_g}\overline{x^2_{III^jI^j}}(t+0.5)
\end{dmath}

With:
\begin{align}
    \overline{x^2_{III^jI^j}}(t) = \frac{1}{N_g(N_g - 1)}\sum_{(i,j) \in I^2 i \neq j} \overline{x_{ii}(t)x_{jj}(t)};
\end{align}

Of course a similar derivation is done for all the products. The interested reader can find these derivations in the appendix.

\subsection{Evolution of the equilibrium average opinion about the agents of group}
\label{sec:biases}
A remarkable feature of the dynamics  that we observe on simulations is that the different opinions $\overline{x_{JI}}(t)$ of groups $J$ about group $I$ tend to be parallel after a while. Following the same idea as in \cite{Deffuant2022}, we now define an equilibrium opinion about group $I$, whose evolution represents this average evolution of the opinions of all groups about group $I$. The equilibrium opinion is obtained by a weighted average o the average self-opinion in group $I$  $\overline{x_{II}}(t)$ and, for all $J \in G$, $\overline{x_{JI}}(t)$ the average opinion of group $J$ about group $I$. The weights are determined so that the first order change of the equilibrium opinion is 0. Therefore, the evolution of this opinion on only due to the second order term which is the core of the average dynamics. Let $e_I(t)$ be this average equilibrium opinion. We have:

\begin{dmath}
   e_I(t)= \frac{1}{1 + S_I(t)}\left(\overline{x_{II}}(t) + \sum_{J \in G}  \frac{(N_g - \delta_{IJ}) \widehat{h_{IJ}}(t)}{\widehat{h_{JI}}(t)} \overline{x_{JI}}(t)\right),
\end{dmath}

where $S_I(t)$ is defined as follows:
\begin{align}
   S_I(t)= \sum_{J \in G}  \frac{(N_g - \delta_{IJ}) \widehat{h_{IJ}}(t)}{\widehat{h_{JI}}(t)}.
\end{align}

Therefore the evolution of $e_I(t)$ is given by:
\begin{dmath}
   e_I(t+1)= e_I(t) + \frac{1}{1 + S_I(t)}\sum_{J \in G} \frac{2(N_g - \delta_{IJ})}{N_c} \left( \overline{h'_{IJ}}(t) (\overline{x^2_{IIJI}}(t) - \overline{x^2_{IIII}}(t)+\overline{x^2_{IIIJ}}(t) - \overline{x^2_{IJJI}}(t)) + \frac{ \widehat{h_{IJ}}(t)}{\widehat{h_{JI}}(t)} \overline{h'_{JI}}(t) (\overline{x^2_{JIJI}}(t) -\overline{x^2_{IIJI}}(t)+ \overline{x^2_{IIJJ}}(t) -\overline{x^2_{JJJI}}(t))\right),
\end{dmath}

We recognise an expression which is very similar to the one found in \cite{Deffuant2022} on the opinions about agents (instead of groups). Similarly, the terms $\overline{h'_{IJ}}(t) (\overline{x^2_{IIJI}}(t) - \overline{x^2_{IIII}}(t)+\overline{x^2_{IIIJ}}(t) - \overline{x^2_{IJJI}}(t))$ are positive and can be interpreted as a "positive bias" about the self-opinions of the groups and the terms $\overline{h'_{JI}}(t) (\overline{x^2_{JIJI}}(t) -\overline{x^2_{IIJI}}(t)+ \overline{x^2_{IIJJ}}(t) -\overline{x^2_{JJJI}}(t))$ are negative and can be interpreted as a "negative bias" of the opinions of groups about other groups and others in their own group.

The tendency of the evolution of the opinion about group $I$ is given by the sign of the sum of the positive bias and negative biases weighted by the coefficient $\frac{\widehat{h_{IJ}}(t)}{\widehat{h_{JI}}(t)}$. This coefficient is large when group $J$ is lower than group $I$ in the group hierarchy (group $J$ has a high opinion of group $I$ and group $I$ a low opinion of group $J$). Therefore, the negative bias is stronger for the groups that are low in the hierarchy. 

When gossip is activated, the equation of $e_i(t+1)$ remains the same except that the following first order term coming from gossip is added:
\begin{align}
  \frac{1}{1 + S_I(t)}\sum_{J \in G}\frac{ \widehat{h_{IJ}}(t)}{\widehat{h_{JI}}(t)}\sum_{P \in G} \frac{2k(N_g - \delta_{IJ})(N_g - \delta_{IP} - \delta_{JP})}{N_t} \widehat{h_{JP}}(t) (\overline{x_{PI}}(t) - \overline{x_{JI}}(t)).
\end{align}
Experimentally, we found that this term remains negligible.

The influence of gossip comes from the term $\overline{x^2_{JIJI}}(t)$ which is increased by gossip, which increases the negative bias, especially for the groups that are low in the hierarchy. These observations are similar to the ones proposed in \cite{Deffuant2022} about the average trajectories of individual opinions.

However, the case of a single group $I$ has no equivalent in the model about individual opinions, because a single agent cannot interact. In this case of a single group $I$, the evolution of the equilibrium opinion about group $I$ is:

\begin{dmath}
   e_I(t+1)= e_I(t) + \frac{\overline{h'_{II}}(t) }{N_a} \left( \overline{x^2_{II^jII^j}}(t) - \overline{x^2_{IIII}}(t) + \overline{x^2_{III^jI^j}}(t) - \overline{x^2_{II^jI^jI}}(t) \right).
\end{dmath}

We observe that, even with gossip, $\overline{x_{IIII}}(t) > \overline{x_{II^jII^j}}(t)$ and $ \overline{x^2_{III^jI^j}}(t) > \overline{x^2_{II^jI^jI}}(t)$ , therefore, the opinion is always increasing (remember that $h'_{II}(t) < 0$). Actually, this corresponds to the case already identified in \cite{Deffuant2022} where the opinions are kept close to each other (by the attraction to the average), leading to an increase of the opinions with or without gossip.

\section{Accuracy of the approximation and explanation of the patterns}

\subsection{Accuracy of the moment approximation}

\subsubsection{Visual representation of opinion trajectories}\label{sec:visualRepresentationAccuracy}

\begin{figure}[!htb]
\begin{spacing}{0.8}
\centering
\begin{tabular}{cc}
   Evolution of $\overline{x_{II}}(t)$ & Evolution of $\overline{x_{JI}}(t)$\\
     \includegraphics[width= 6 cm]{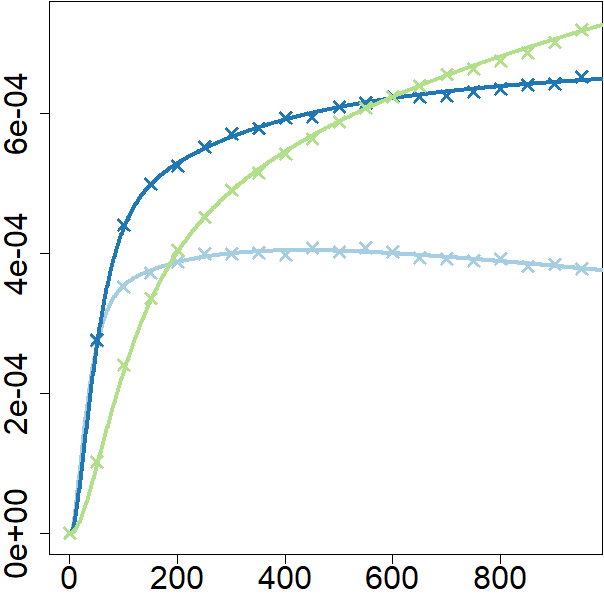} & \includegraphics[width= 6 cm]{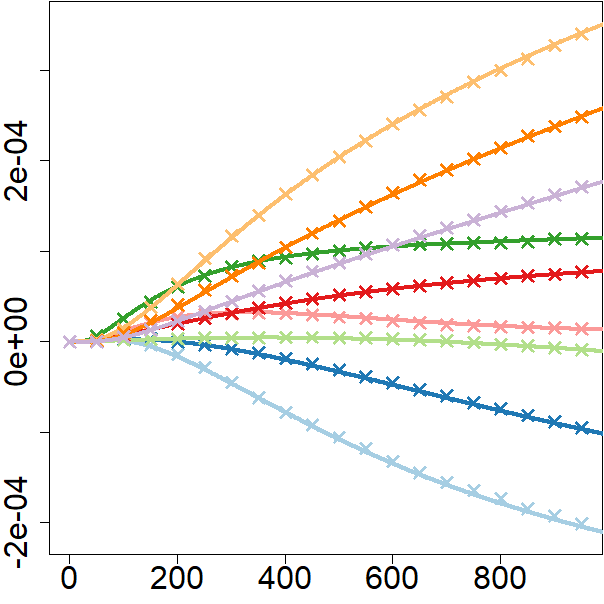} \\
    $t$ & $t$
\end{tabular}
\end{spacing}
   \caption{Examples of evolution of average group opinions for 3 groups of 10 agents. The lines are obtained with the moment approximation and the crosses by averaging the results of 500 thousands simulations. Noise $\delta = 0.05$, influence parameter $\sigma = 0.3$, attraction to group average $\mu = 0.995$, gossip $k=2$. See main text for details. }
	\label{fig:accuracy}
\end{figure}

Figure \ref{fig:accuracy} shows the evolution of average self-opinions of group $I$ agents $\overline{x_{II}}(t)$ (left panel) and average opinions of group $I$ about group $J$ agents $\overline{x_{JI}}(t)$ (right panel) for 3 groups of 10 agents. Initially, all the agents have the opinion -0.5 about agents of group 0, the opinion 0 about agents of group 1, opinion 0.5 about agents of group 2. The panels show the average value over 500,000 simulations (crosses) and the moment approximation. The approximation appears very accurate. It is also noticeable that the average opinion about some groups increases while the opinion about other groups decreases.

\subsubsection{RRMSE of the moment approximation.}

We use the root of the relative mean squared error (RRMSE) to evaluate the difference between the moment of approximation and the average results over 500,000 simulations. The setting for the experiment is the same as in the previous subsection (\ref{sec:visualRepresentationAccuracy}).

Keeping the notation $\overline{x_{II}}(t)$ for the moment approximation, using $\widehat{x_{II}}(t)$ for the average of the simulations, and 
 $\mathcal{E}\left(\overline{x_{II}}(1,\dots,T)\right)$ the RRMSE  $\overline{x_{II}}(t)$ for $t \in [1, T]$, is defined as:

\begin{align} \label{eq:RRMSE}
    \mathcal{E}\left(\overline{x_{II}}(1,\dots,T)\right) = \frac{\sqrt{\frac{1}{T}\left(\sum_{t = 1}^{T}\left( \overline{x_{II}}(t) - \widehat{x_{II}}(t)\right)^2\right)}}{\sum_{t = 1}^{T} |\widehat{x_{II}}(t)|}.
\end{align}
We define the RRMSE for $x_{JI}$ and the second moment variables similarly.
 
Figure \ref{fig:RRMSE} shows the average RRMSE computed for different moment approximations, with and without gossip. The RRMSE is lower than $5.10^{-4}$ without gossip, and on average lower than $5.10^{-3}$ with gossip. The error is higher for the average group opinions, especially for $\overline{x_{JI}}$, than for the average products of group opinions. The errors decrease with maximum time $T$, probably because the absolute values of $\overline{x_{IJ}}$ increase more than the approximation errors with $t$.

Overall these tests suggest that the moment approximation provides reliable values of the evolution of the group opinions in the model, avoiding to average the results of millions of simulations.

\begin{figure}[!htb]\label{fig:RRMSE}
    \centering
    \begin{tabular}{cc}
       No gossip ($k=0$) & Gossip ($k=2$)\\
         \includegraphics[width= 6 cm]{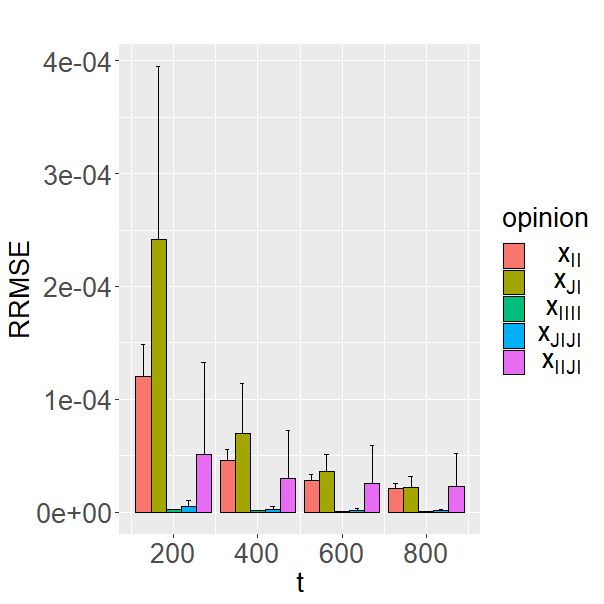} & \includegraphics[width= 6 cm]{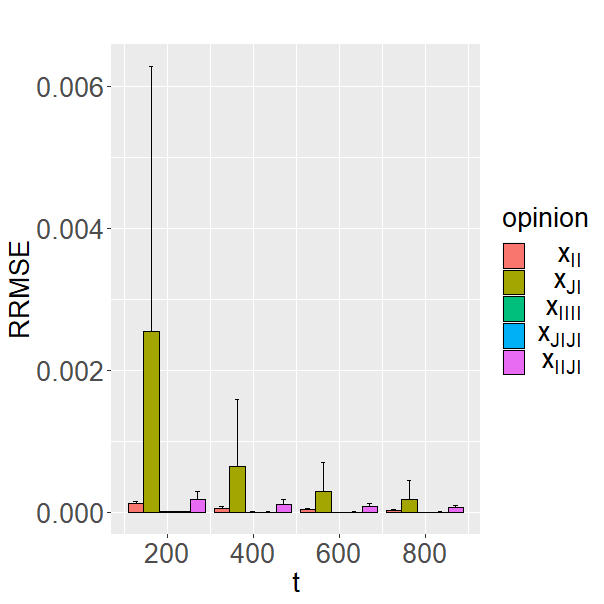} 
    \end{tabular}
   \caption{Average RRMSE of the approximation (defined by equation \ref{eq:RRMSE}). Left panel: no gossip ($k=0$). Left panel: gossip ($k=2$) Bottom panels average for all second moment variables. The RRMSE is computed on the interval $[1,t]$, $t$ being defined on the horizontal axis. The error bars show the standard deviations in the considered set of variables ($N_g$ variables for $x_{II}$, $N_g^2$ for $x_{JI}$, etc...). Number of group $ng=3$, number of agent per group $N_g=10$, noise $\delta = 0.05$, influence parameter $\sigma = 0.3$, attraction to group average $\mu = 0.995$.}
    \label{fig:RRMSE}
\end{figure}

\subsubsection{The effect of inequalities explains the patterns.}

In order to explain the patterns presented in section \ref{sec:patterns}, we compute the average tendency of the evolution of the group opinions starting from more or less strong hierarchical differences between groups, using the moment approximation. We start with 2 groups and then we consider 3 groups. We compute the trend of the evolution of the equilibrium opinion about group $I$ at $t = 1000$ as $e_I(1000) - e_I(999)$, because, experimentally, after 1000 steps, the trend stabilises. Hence this gives the tendency of the evolution of the opinions about the group for a while after (about several thousand steps). In both cases, the parameter values are as follows: number of agents in a group $N_g = 10$, noise $\delta = 0.05$, influence parameter $\sigma = 0.3$, attraction to group average $\mu = 0.995$, gossip parameter $k=5$. 

\begin{figure}[!htb]
\centering
\begin{tabular}{cc}
    group 0 (high status) & group 1 (low status) \\
    \includegraphics[width= 6 cm]{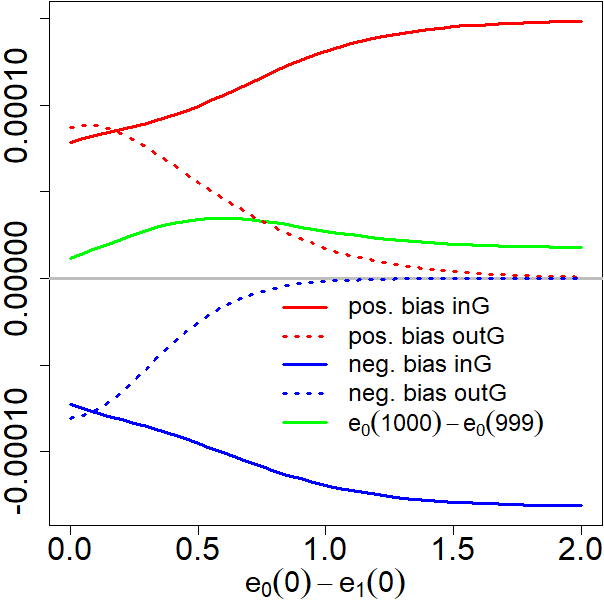}  & \includegraphics[width= 6 cm]{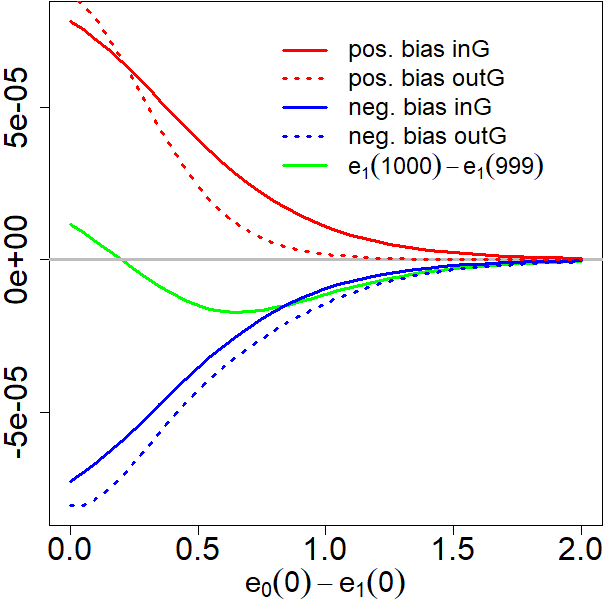}
\end{tabular}
    \caption{2 groups. Biases and equilibrium trajectory change  at $t = 1000$ computed by the moment approximation, for different initial differences of opinions about groups. The left panel considers the group of high status and the right panel the group of low status. The red curves represent the positive biases and the blue curves the negative biases. The solid lines represent the contribution of in-group interactions and the dotted lines the contribution of out-group interactions in the biases. The green lines represent the weighted sum of the biases, giving the change of the equilibrium opinion at $t = 1000$.}
	\label{fig:modelEvol2g}
\end{figure}

In the case of two groups (group 0 and group 1), all the agents are initialised with the same opinion $e_0(0)$ about agents of group 0 and the same opinion $e_1(0)$ about the agents of group 1. Moreover, group 0 is assumed of higher status: $e_0(0) > e_1(0)$. The the difference $e_0(0) - e_1(0)$ varies from 0.01 to 2, regularly with intervals of 0.01. For each value, we compute the moment approximation for 1000 steps.

Figure \ref{fig:modelEvol2g} shows the values of the positive and negative biases as defined in section \ref{sec:biases} in the case of 2 groups (0 and 1) and their weighted sum giving the change of equilibrium opinion $e_I(1000) - e_I(999)$ about the considered group. The horizontal axis represents the initial difference of opinions about the two groups $e_0(0) - e_1(1)$. The graphs show that the biases from in-group interactions are stronger than the biases from out-group interactions for the group of high status (left panel). The in-group and out-group biases are of similar amplitude for the group of low status (left panel). Not that the difference between the positive and the negative biases is larger for the out-group interactions. Hence, the out-group interactions are mainly responsible for the tendency of the evolution of the equilibrium opinion about the group.

Note that the change of equilibrium opinion about the group of high status (green curve) is always positive, hence this opinion is increasing whatever the initial opinion difference.
On the contrary, when the difference $e_0(0) - e_1(0)$ reaches a threshold (about 0.4), the change of equilibrium opinion about the group of low status (green curve) becomes negative indicating that the opinions about this group decrease. 

These observations explain the shape of the evolution of the opinions about the groups shown on figure \ref{fig:opMatrix}. Indeed, first during a few thousand steps, there is no clear dominance of one group and the opinions about both fluctuate around 0. Then, because of random fluctuations, the yellow group dominates for a while and the opinion about it grows steadily. Initially the opinion about the orange group increases as well, but less rapidly and the difference of status increases, until reaching the threshold for which the opinion about the orange group decreases. Note that this decreasing tendency attenuates when the status gap increases, which explains why the opinion about the orange group can increase by fluctuations and then decrease again.

\begin{figure}[!htb]
\centering
	\includegraphics[width= 6 cm]{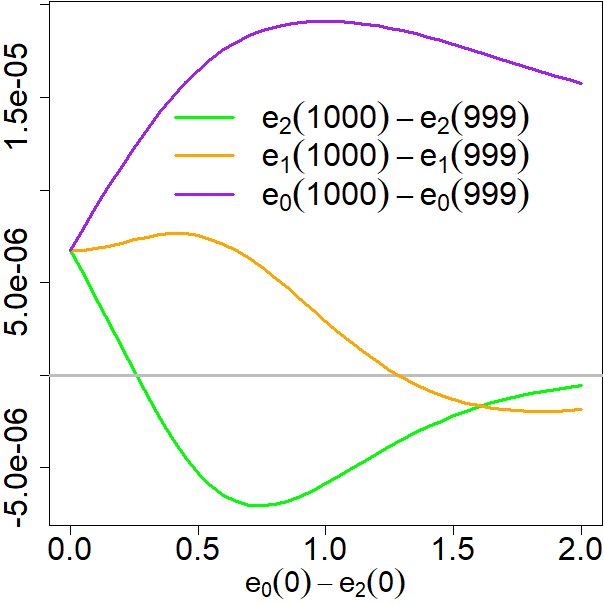}
    \caption{3 groups. Slope of equilibrium trajectories at $t = 1000$ ( $100 \times e_0(1000)-e_0(999)$) computed by the moment approximation, for 3 groups of 10 agents, for different initial gap of reputations. The 3 groups (0, 1, and 2), with group 0 being the highest group, group 2 the lowest group, and group 1 initially in the middle.Noise $\delta = 0.05$, influence parameter $\sigma = 0.3$, attraction to group average $\mu = 0.995$, gossip parameter $k=5$.}
	\label{fig:modelEvol3g}
\end{figure}

Figure \ref{fig:modelEvol3g} considers the case of 3 groups and shows the change of equilibrium opinion after 1000 random interactions for each group. The curves are the values $e_I(1000) - e_I(999)$ for $I \in \{0, 1, 2\}$, for different initial differences $e_0(0) - e_2(0)$,assuming that the initial opinion about each group $I$ is shared by all the agents and that the hierarchy is initially regular: $e_1(0) = \frac{1}{2}(e_0(0) + e_1(0))$. We note that $e_0(t)$ decreases when $e_0(0)$ is significantly lower than $e_2(0)$ (approximately $e_0(0) - e_2(0) > 0.3$) and $e_1(t)$ decreases (slightly) only when $e_0(0)$ is much lower than $e_2(0)$ (approximately $e_0(0) - e_2(0) > 1.5$). Therefore, when the difference between the two extreme groups becomes high, the opinions about the intermediate group tend to decrease.

This analysis also provides with some keys to understand the evolution of opinions for 3 groups displayed on figure \ref{fig:opMatrix}. Indeed, the opinion about the group of lowest status (green) starts decreasing when the difference with the group of highest status is large enough (around 0.3). The opinion about the group of intermediate status initially grows with the opinion about the group of highest status, until the difference between the two extreme groups reaches a value around 1.5, and then the opinion about the group of intermediate status starts to decrease.

\section{Discussion}

In this model, the dominance of a group comes from random differences that tend to increase because the highest in the hierarchy tends to get higher while the lowest tends to get lower.
The process is very similar to the emergence of a hierarchy of agents in the model without group. We should investigate the model with larger numbers of groups to confirm if the order of groups is unstable at the top of the hierarchy like it is for individuals without groups.
In any case, the main conclusion of this work is that the average self-opinion of the agents of a group has a positive bias, like the agents have a positive bias on their self-opinion, and the average opinion of the agents of a group about the agents of another group has a negative bias, like the agents have a negative bias on their opinion about other agents. Similarly also to the agents without groups, the relative effect of the negative bias is stronger for the groups which are low in the hierarchy. Therefore, the model suggests that, even if the differences between groups are purely anecdotal and have no connections with the intrinsic qualities of the agents, any society tends to establish a group hierarchy, purely because it tends to increase advantages or disadvantages that are initially randomly established.

Moreover, it should be stressed that the hierarchy emerges only when agents gossip. Indeed, without gossip, all the opinions about all groups grow together. This echoes the results from the model without groups, in which gossiping tends to spread the hierarchy by decreasing the opinions about the agents of lower status. This effect is even more contrasted with groups, because the opinions tend to be close to each other within each group, which reinforces the positive bias within groups.

The limitations mentioned in \cite{Deffuant2022} of course remain valid for this work. Mainly, the model does not pretend to be realist. On the contrary, it is simplified on purpose, in order to make the effect of some hypotheses clearer. Moreover, in the model with groups, the process by which the opinions in a group are attracted to their average is probably more artificial than the others because it is unlikely that the agents have access to the exact average opinion of the group about the agents of another group. It is indeed designed in order to facilitate the derivation of the moment approximation. Nevertheless, it roughly represents the effect of a shared prejudice. The conclusions drawn with this simplified process are likely to remain valid when replacing it by a better grounded one in which this shared prejudice varies slightly among the agents of the group.

\section{Acknowledgement}

This work has been partly supported by the Agence Nationale de la Recherche through the ToRealSim project (ORA call).

\section{Additional information}
The code of the model is available at: \url{https://www.comses.net/codebases/197f157c-ea03-49fb-bd5c-db3a943f8a99/releases/1.0.0/}.

\bibliographystyle{splncs04}
\bibliography{references}

\begin{thebibliography}{10}
\providecommand{\url}[1]{\texttt{#1}}
\providecommand{\urlprefix}{URL }
\providecommand{\doi}[1]{https://doi.org/#1}

\bibitem{Bagnoli2007}
Bagnoli, F., Carletti, T., Fanelli, D., Guarino, A., Guazzini, A.: Dynamical
  affinity in opinion dynamics modeling. Physical Review E  \textbf{76} (2007)

\bibitem{Bonabeau1996}
Bonabeau, E., Th\'eraulaz, G., Deneubourg, J.L.: Mathematical model of
  self-organizing hierarchies in animal societies. Bulletin of Mathematical
  Biology  \textbf{58}(4),  661--717 (1996)

\bibitem{Carletti2011}
Carletti, T., Fanelli, D., Simone, R.: Emerging structures in social networks
  guided by opinions. Advances in Complex Systems  \textbf{14} (2011)

\bibitem{Deffuant2006a}
Deffuant, G.: Comparing extremism propagation patterns in continuous opinion
  models. Journal of Artificial Societies and Social Simulation  \textbf{9}(3)
  (2006)

\bibitem{Deffuant2018}
Deffuant, G., Bertazzi, I., Huet, S.: The dark side of gossips: Hints from a
  simple opinions dynamics model:. Advances in Complex Systems  \textbf{21}
  (2018)

\bibitem{Deffuant2013}
Deffuant, G., Carletti, T., Huet, S.: The leviathan model: Absolute dominance,
  generalised distrust and other patterns emerging from combining vanity with
  opinion propagation. Journal of Artificial Societies and Social Simulation
  \textbf{16}(23) (2013)

\bibitem{Deffuant2000}
Deffuant, G., Neau, D., Amblard, F., Weisbuch, G.: Mixing beliefs among
  interacting agents. Advances in Complex Systems  \textbf{3},  87--98 (2000)

\bibitem{Deffuant2022}
Deffuant, G., Roubin, T.: Do interactions among unequal agents undermine those
  of low status? Physica A: Statistical Mechanics and its Applications
  \textbf{592},  126780 (2022).
  \doi{https://doi.org/10.1016/j.physa.2021.126780},
  \url{https://www.sciencedirect.com/science/article/pii/S0378437121009626}

\bibitem{Flache2017}
Flache, A., Maes, M., Feliciani, T., Chattoe-Brown, E., Deffuant, G., Huet, S.,
  Lorenz, J.: Models of social influence: Towards the next frontiers. Journal
  of Artificial Societies and Social Simulation  \textbf{20}(4) (2017)

\bibitem{Flache2018}
Flache, A.: About renegades and outgroup haters: Modeling the link between
  social influence and intergroup attitudes. Advances in Complex Systems
  \textbf{21}(06n07) (2018)

\bibitem{Flache2011}
Flache, A., Macy, M.: Small worlds and cultural polarization. The Journal of
  Mathematical Sociology  \textbf{35}(1-3),  146--176 (2011)

\bibitem{French1956}
French, J.: A formal theory of social power. Psychological Review
  \textbf{63}(3),  181--194 (1956)

\bibitem{Galam2002}
Galam, S.: Minority opinion spreading in random geometry. The European Physical
  Journal B  \textbf{25}(4),  403--406 (2002)

\bibitem{Hegselmann2002}
Hegselmann, R., Krause, U.: Opinion dynamics and bounded confidence: Models,
  analysis and simulation. Journal of Artificial Societies and Social
  Simulation  \textbf{5}(3) (2002)

\bibitem{Holyst2001}
Holyst, J., Kacperski, K., Schweitzer, F.: Social impact models of opinion
  dynamics. Annual review of Computational Physics IX pp. 253--273 (2001)

\bibitem{Latane1981}
Latan{\'e}, B.: The psychology of social impact. American Psychologist
  \textbf{36},  343--356 (1981)

\bibitem{Law2000}
Law, R., Dieckmann, U.: Moment approximations of individual-based models. In:
  Dieckmann, U. (ed.) The Geometry of Ecological Interactions: Simplifying
  Spatial Complexity., pp. 252--269. Cambridge University Press (2000)

\end{thebibliography}

\section{Appendix: equations of the evolution of average products}

This appendix provides the equations ruling the approximate average evolution of products of average group opinions.

\subsubsection*{Expression of $x^2_{IIII}(t+0.5)$}We consider the product $x^2_{IIII}(t)$.

\begin{dmath}
    \overline{x^2_{ii}}(t+0.5)  =  \overline{x^2_{ii}}(t) + \frac{2}{N_c} \sum_{J \in G} \sum_{j \in J, j \neq i} \left((\widehat{h^2_{IJ}}(t)-2\widehat{h_{IJ}}(t))\overline{x^2_{ii}}(t) + \widehat{h_{IJ}}(t)^2\overline{x^2_{ji}}(t) + 2 (1-\widehat{h_{IJ}}(t))\widehat{h_{IJ}}(t)\overline{x_{ii}(t)x_{ji}(t)} + \widehat{h_{IJ}}(t)^2 \frac{\delta^2}{3} \right).
\end{dmath}

Therefore:
\begin{dmath}
    \overline{x^2_{IIII}}(t+0.5)  = \overline{x^2_{IIII}}(t) + \sum_{J \in G}\frac{2(N_g - \delta_{IJ})}{N_c} \left((\widehat{h^2_{IJ}}(t)-2\widehat{h_{IJ}}(t))\overline{x^2_{IIII}}(t) + \widehat{h_{IJ}}(t)^2\overline{x^2_{JIJI}}(t) + 2 (1-\widehat{h_{IJ}}(t))\widehat{h_{IJ}}(t)\overline{x^2_{IIJI}(t)} + \widehat{h_{IJ}}(t)^2 \frac{\delta^2}{3} \right).
\end{dmath}
 Note that the expression $\frac{\delta^2}{3}$ appearing in the previous equation comes from the average of the squared noise $\overline{\theta^2}(t) = \frac{\delta^2}{3}$ (see \cite{Deffuant2022} for details).

\subsubsection*{Expression of $x^2_{IIII}(t+1)$}Now, we express how these quantities change because of the process of attraction of the average opinion.

\begin{dmath}
   \overline{x_{ii}^2}(t+1)  =  \left(\mu\overline{x_{ii}}(t+0.5)+ \frac{1 - \mu}{N_g}\sum_{j \in I}\overline{x_{jj}}(t+0.5)\right)^2
\end{dmath}

\begin{dmath}
   \overline{x_{ii}^2}(t+1)  =  \mu^2\overline{x_{ii}^2}(t+0.5)+ \frac{2\mu(1 - \mu)}{N_g}\overline{x_{ii}}(t+0.5)\sum_{j \in I}\overline{x_{jj}}(t+0.5)
   + \frac{(1 - \mu)^2}{N_g^2}\sum_{j \in I}\sum_{p \in I}\overline{x_{jj}(t+0.5)x_{pp}(t+0.5)}
\end{dmath}

\begin{dmath}
   \frac{1}{N_g}\sum_{i \in I} \overline{x_{ii}^2}(t+1)  =  \frac{\mu^2}{N_g}\sum_{i \in I}\overline{x_{ii}^2}(t+0.5)+ \frac{2\mu(1 - \mu)}{N_g^2}\sum_{i \in I}\sum_{j \in I}\overline{x_{ii}(t+0.5)x_{jj}(t+0.5)}
   + \frac{(1 - \mu)^2}{N_g^2}\sum_{j \in I}\sum_{p \in I}\overline{x_{jj}(t+0.5)x_{pp}(t+0.5)}
\end{dmath}

Therefore:

\begin{dmath}
   \overline{x^2_{IIII}}(t+1)  = \left(\mu^2 + \frac{1-\mu^2}{N_g} \right)  \overline{x^2_{IIII}}(t+0.5)+ \frac{(1-\mu^2)(N_g-1)}{N_g}\overline{x^2_{III^jI^j}}(t+0.5)
\end{dmath}

With:
\begin{align}
    \overline{x^2_{III^jI^j}}(t) = \frac{1}{N_g(N_g - 1)}\sum_{(i,j) \in I^2 i \neq j} \overline{x_{ii}(t)x_{jj}(t)};
\end{align}

The other products follow the same principles

\subsubsection*{Expression of $x^2_{IIIJ}(t+0.5):$}

This product is null for $I \neq J$.
For $(i,j) \in I \times J, i \neq j$:
\begin{dmath}
    \overline{x_{ii}(t+0.5)x_{ij}(t+0.5)}  =  \overline{x_{ii}(t)x_{ij}(t)} + \frac{2}{N_c} \sum_{P \in G} \sum_{p \in P, p \notin (i,j)} \left(\widehat{h_{IP}}(t)(\overline{x_{ij}(t)x_{pi}(t)} - \overline{x_{ii}(t)x_{ij}(t)})\right)  +  \frac{2}{N_c} \left(\widehat{h_{IJ}}(t)(\overline{x_{ji}(t)x_{ij}(t)} - \overline{x_{ii}(t)x_{ij}(t)}) + \widehat{h_{IJ}}(t)( \overline{x_{ii}(t)x_{jj}(t)} - \overline{x_{ii}(t)x_{ij}(t)} ) + \widehat{h_{IJ}}^2(t)( \overline{x_{jj}(t)x_{ji}(t)} +  \overline{x_{ii}(t)x_{ij}(t)} -  \overline{x_{ii}(t)x_{jj}(t)} -  \overline{x_{ij}(t)x_{ji}(t)})\right) +\frac{2k}{N_T}\sum_{P \in G} \sum_{p \in P, p \notin \{i,j\}} \widehat{h_{IP}}(t)\left( \overline{x_{ii}(t)x_{pj}(t)}
    -\overline{x_{ii}(t)x_{ij}(t)}\right).
\end{dmath}

Summing up on all $(i,j) \in I \times J, i \neq j$ and dividing by $N_g (N_g - \delta_{IJ})$, we get:

\begin{dmath}
    \overline{x^2_{IIIJ}(t+0.5)}  =  \overline{x^2_{IIIJ}(t)} +   \frac{2}{N_c} \sum_{P \in G} \left( (N_g - \delta_{JP} - \delta_{IP})\widehat{h_{IP}}(t)(\overline{x^2_{IJPI}(t)} - \overline{x^2_{IIIJ}(t)}) \right)
    +  \frac{2}{N_c} \left(\widehat{h_{IJ}}(t)( \overline{x^2_{JIIJ}(t)} + \overline{x^2_{IIJJ}(t)} - 2\overline{x^2_{IIIJ}(t)}) + \widehat{h_{IJ}}^2(t)( \overline{x^2_{JJJI}(t)} - \overline{x^2_{JIIJ}(t)} - \overline{x^2_{IIJJ}(t)} + \overline{x^2_{IIIJ}(t)} ) \right)
    +\frac{2k}{N_T}\sum_{P \in G} (N_g - \delta_{IP} - \delta_{JP})\widehat{h_{IP}}(t)\left(  \overline{x^2_{IIPJ}(t)} -  \overline{x^2_{IIIJ}(t)} \right).
\end{dmath}

\subsubsection*{Expression of $x^2_{IIIJ}(t+1)$:}

With the same derivation, for $I = J$, the only case where this product is not null, we get:
\begin{dmath}
   \overline{x^2_{IIII^j}(t+1)}   = \left( \mu^2+ \frac{(1 - \mu^2)}{N_g}\right)\overline{x^2_{IIII^j}(t+0.5)} + \frac{1 - \mu^2}{N_g} \overline{x^2_{III^jI}(t+0.5)} + \frac{(1 - \mu^2)(N_g - 2)}{N_g} \overline{x^2_{III^jI^q}(t+0.5)}.
\end{dmath}

\subsubsection*{Expression of $x^2_{IIJI}(t+0.5):$} 

For $(i,j) \in I \times J, i \neq j$:
\begin{dmath}
    \overline{x_{ii}(t+0.5)x_{ji}(t+0.5)}  =  \overline{x_{ii}(t)x_{ji}(t)} + \frac{2}{N_c} \sum_{P \in G} \sum_{p \in P, p \notin (i,j)} \left(\widehat{h_{IP}}(t)(\overline{x_{ji}(t)x_{pi}(t)} - \overline{x_{ii}(t)x_{ji}(t)})\right)  +  \frac{2}{N_c} \left(\widehat{h_{IJ}}(t)(\overline{x_{ji}(t)x_{ji}(t)} - \overline{x_{ii}(t)x_{ji}(t)}) + \widehat{h_{JI}}(t)( \overline{x_{ii}(t)x_{ii}(t)} - \overline{x_{ii}(t)x_{ji}(t)} ) + \widehat{h_{IJ}}(t)\widehat{h_{JI}}(t)( 2\overline{x_{ji}(t)x_{ii}(t)} -  \overline{x_{ii}(t)x_{ii}(t)} -  \overline{x_{ji}(t)x_{ji}(t)})\right) +\frac{2k}{N_T}\sum_{P \in G} \sum_{p \in P, p \notin \{i,j\}} \widehat{h_{JP}}(t)\left( \overline{x_{ii}(t)x_{pi}(t)}
    -\overline{x_{ii}(t)x_{ji}(t)}\right).
\end{dmath}

Summing up on all $(i,j) \in I \times J, i \neq j$ and dividing by $N_g (N_g - \delta_{IJ})$, we get:

\begin{dmath}
    \overline{x^2_{IIJI}(t+0.5)}  =  \overline{x^2_{IIJI}(t)} +   \frac{2}{N_c} \sum_{P \in G} \left( (N_g - \delta_{IP})\widehat{h_{IP}}(t)(\overline{x^2_{JIPI}(t)} - \overline{x^2_{IIJI}(t)}) \right)+ \frac{2}{N_c} \left(\widehat{h_{IJ}}(t)( \overline{x^2_{IJJI}(t)} -  \overline{x^2_{IIJI}(t)}) + \widehat{h_{JI}}(t)( \overline{x^2_{IIII}(t)} -  \overline{x^2_{IIJI}(t)}) + \widehat{h_{IJ}}(t)\widehat{h_{JI}}(t)( 2\overline{x^2_{IIJI}(t)}  - \overline{x^2_{IIII}(t)} - \overline{x^2_{JIJI}(t)}) \right) +\frac{2k}{N_T}\sum_{P \in G} (N_g - \delta_{IP} - \delta_{JP}) \widehat{h_{JP}}(t)\left(  \overline{x^2_{IIPI}(t)} -  \overline{x^2_{IIJI}(t)} \right).
\end{dmath}

\subsubsection*{Expression of $x^2_{IIJI}(t+1)$:}

For $(i,j) \in I \times J, i \neq j$: 
\begin{dmath}
   \overline{x_{ii}(t+1)x_{ji}(t+1)}   =  \left(\mu\overline{x_{ii}}(t+0.5)+ \frac{1 - \mu}{N_g}\sum_{q \in I}\overline{x_{pp}}(t+0.5)\right)\left(\mu\overline{x_{ji}}(t+0.5)+ \frac{1 - \mu}{N_g(N_g - 1)}\sum_{(r,s) \in I\times J, r \neq s}\overline{x_{rs}}(t+0.5)\right)
\end{dmath}

\begin{dmath}
   \overline{x_{ii}(t+1)x_{ji}(t+1)}   =  \mu^2\overline{x_{ii}(t+0.5)x_{ji}(t+0.5)}+ \frac{\mu(1 - \mu)}{N_g}\sum_{p \in I,}\overline{x_{pp}(t+0.5)x_{ji}(t+0.5)}  + \frac{\mu(1 - \mu)}{N_g(N_g - \delta_{IJ})}\sum_{(r,s) \in J \times I, r \neq s}\overline{x_{ii}(t+0.5)x_{rs}(t+0.5)}+  \frac{(1 - \mu)^2}{N^2_g(N_g - \delta_{IJ})}\sum_{(p,s,r) \in I^2 \times J, r \neq s}\overline{x_{pp}(t+0.5)x_{rs}(t+0.5)}.
\end{dmath}

Summing up on all $(i,j) \in I \times J, i \neq j$ and dividing by $N_g (N_g - \delta_{IJ})$, we get:

\begin{dmath}
   \overline{x^2_{IIJI}(t+1)}   = \left( \mu^2+ \frac{1 - \mu^2}{N_g}\right)\overline{x^2_{IIJI}(t+0.5)} + \frac{(1 - \mu^2)(N_g-1-\delta_{IJ})}{N_g} \overline{x^2_{IIJI^q}(t+0.5)} + \frac{\delta_{IJ}(1 - \mu^2)}{N_g} \overline{x^2_{IIII^q}(t+0.5)}.
\end{dmath}

Where:
\begin{dmath}
    \overline{x^2_{IIJI^q}}(t) = \frac{1}{N_g(N_g - \delta_{IJ})(N_g - \delta_{IJ}-1)}\sum_{(i,q,j) \in I^2 \times J, \mbox{ all distinct}} \overline{x_{ii}(t)x_{jq}(t)}.
\end{dmath}

\subsubsection*{Expression of $x^2_{IIJJ}(t+0.5):$}
This product is null for $I \neq J$.
For $(i,j) \in I \times J, i \neq j$:

\begin{dmath}
    \overline{x_{ii}(t+0.5)x_{jj}(t+0.5)}  =  \overline{x_{ii}(t)x_{jj}(t)} + \frac{2}{N_c} \sum_{P \in G} \sum_{p \in P, p \neq i} \left(\widehat{h_{IP}}(t)(\overline{x_{jj}(t)x_{pi}(t)} - \overline{x_{jj}(t)x_{ii}(t)})\right) + \frac{2}{N_c} \sum_{P \in G} \sum_{p \in P, p \neq j} \left(\widehat{h_{JP}}(t)(\overline{x_{ii}(t)x_{pj}(t)} - \overline{x_{jj}(t)x_{ii}(t)})\right) +  \frac{2}{N_c} \left(\widehat{h_{IJ}}(t)( \overline{x_{jj}(t)x_{ji}(t)} -  \overline{x_{ii}(t)x_{jj}(t)}) + \widehat{h_{JI}}(t) (\overline{x_{ii}(t)x_{ij}(t)} -  \overline{x_{ii}(t)x_{jj}(t)}) + \widehat{h_{IJ}}(t)\widehat{h_{JI}}(t)( \overline{x_{jj}(t)x_{ii}(t)} +  \overline{x_{ji}(t)x_{ij}(t)} -  \overline{x_{ii}(t)x_{ij}(t)} -  \overline{x_{jj}(t)x_{ji}(t)}) \right).
\end{dmath}

Summing up on all $(i,j) \in I \times J, i \neq j$ and dividing by $N_g (N_g - \delta_{IJ})$, we get:

\begin{dmath}
    \overline{x^2_{IIJJ}(t+0.5)}  =  \overline{x^2_{IIJJ}(t)} + \frac{2}{N_c} \sum_{P \in G} \left( (N_g - \delta_{IP} - \delta_{JP})\widehat{h_{IP}}(t)(\overline{x^2_{JJPI}(t)} - \overline{x^2_{IIJJ}(t)})\right) + \frac{2}{N_c} \sum_{P \in G} \left( (N_g - \delta_{IP} - \delta_{JP})\widehat{h_{JP}}(t)(\overline{x^2_{IIPJ}(t)} - \overline{x^2_{IIJJ}(t)})\right) +  \frac{2}{N_c} \left(\widehat{h_{IJ}}(t)(\overline{x^2_{JJJI}(t)} -  \overline{x^2_{IIJJ}(t)}) + \widehat{h_{JI}}(t)(\overline{x^2_{IIIJ}(t)}  -  \overline{x^2_{IIJJ}(t)}) + \widehat{h_{IJ}}(t)\widehat{h_{JI}}(t)( \overline{x^2_{IIJJ}(t)} +  \overline{x^2_{IJJI}(t)} -  \overline{x^2_{IIIJ}(t)} -  \overline{x^2_{JJJI}(t)}) \right).
\end{dmath}

\subsubsection*{Expression of $x^2_{IIJJ}(t+1)$:}

For $(i,j) \in I \times J, i \neq j$:

\begin{dmath}
   \overline{x_{ii}(t+1)x_{jj}(t+1)}   =  \left(\mu\overline{x_{ii}}(t+0.5)+ \frac{1 - \mu}{N_g}\sum_{p \in I}\overline{x_{pp}}(t+0.5)\right)\left(\mu\overline{x_{jj}}(t+0.5)+ \frac{1 - \mu}{N_g}\sum_{q \in J}\overline{x_{qq}}(t+0.5)\right)
\end{dmath}

\begin{dmath}
   \overline{x_{ii}(t+1)x_{jj}(t+1)}   =  \mu^2\overline{x_{ii}(t+0.5)x_{jj}(t+0.5)} + \frac{\mu(1 - \mu)}{N_g}\sum_{p \in I}\overline{x_{pp}(t+0.5)x_{jj}(t+0.5)} + \frac{\mu(1 - \mu)}{N_g}\sum_{q \in J}\overline{x_{ii}(t+0.5)x_{qq}(t+0.5)} + \frac{(1 - \mu)^2}{N^2_g}\sum_{(p,q) \in I \times J}\overline{x_{pp}(t+0.5)x_{qq}(t+0.5)}
\end{dmath}

Summing up on all $(i,j) \in I^2, i \neq j$ and dividing by $N_g (N_g - \delta_{IJ})$, we get:

For $I = J$, the only case this product is not null:
\begin{dmath}
   \overline{x^2_{III^jI^j}(t+1)}   = \left(\mu^2 - \frac{(1 - \mu^2)(N_g - 1)}{N_g} \right)\overline{x^2_{III^jI^j}(t+0.5)} + \frac{(1 - \mu^2)}{N_g} \overline{x^2_{IIII}(t+0.5)}.
\end{dmath}

\subsubsection*{Expression of $x^2_{IJJI}(t+0.5):$}
This product is null for $I \neq J$.
For $(i,j) \in I \times J, i \neq j$:
\begin{dmath}
    \overline{x_{ij}(t+0.5)x_{ji}(t+0.5)}  =  \overline{x_{ij}(t)x_{ji}(t)}  +  \frac{2}{N_c}\left(\widehat{h_{IJ}}(t)(\overline{x_{jj}(t)x_{ji}(t)} - \overline{x_{ij}(t)x_{ji}(t)}) + \widehat{h_{JI}}(t)(\overline{x_{ii}(t)x_{ij}(t)} - \overline{x_{ij}(t)x_{ji}(t)} ) + \widehat{h_{IJ}}(t)\widehat{h_{JI}}(t)( \overline{x_{jj}(t)x_{ii}(t)} +  \overline{x_{ji}(t)x_{ij}(t)} -  \overline{x_{ii}(t)x_{ij}(t)} -  \overline{x_{jj}(t)x_{ji}(t)}) \right) +\frac{2k}{N_T}\sum_{P \in G} \sum_{p \in P, p \notin \{i,j\}}\left( \widehat{h_{JP}}(t) (\overline{x_{ij}(t)x_{pi}(t)}
    -\overline{x_{ij}(t)x_{ji}(t)}) +\widehat{h_{IP}}(t)(\overline{x_{ji}(t)x_{pj}(t)}
    -\overline{x_{ji}(t)x_{ij}(t)}) \right).
\end{dmath}

Summing up on all $(i,j) \in I \times J, i \neq j$ and dividing by $N_g (N_g - \delta_{IJ})$, we get:

\begin{dmath}
    \overline{x^2_{IJJI}(t+0.5)}  =  \overline{x^2_{JJJI}(t)} +   \frac{2}{N_c} \left(\widehat{h_{IJ}}(t)( \overline{x^2_{JJJI}(t)} - \overline{x^2_{IJJI}(t)}) + \widehat{h_{JI}}(t)( \overline{x^2_{IIIJ}(t)} - \overline{x^2_{IJJI}(t)})  + \widehat{h_{II}}(t)\widehat{h_{JI}}(t)( \overline{x^2_{IIJJ}(t)} +  \overline{x^2_{IJJI}(t)} -  \overline{x^2_{JJJI}(t)} -  \overline{x^2_{IIIJ}(t)}) \right) + \frac{2k}{N_T}\sum_{P \in G} (N_g - \delta_{IP} - \delta_{JP})\left( \widehat{h_{JP}}(t) (\overline{x^2_{IJPI}(t)} - \overline{x^2_{IJJI}(t)}) +  \widehat{h_{IP}}(t) (\overline{x^2_{JIPJ}(t)} - \overline{x^2_{IJJI}(t)}) \right).
\end{dmath}

\subsubsection*{Expression of $x^2_{IJJI}(t+1)$:}

For $(i,j) \in I \times J, i \neq j$: 
\begin{dmath}
   \overline{x_{ij}(t+1)x_{ji}(t+1)}   =  \left(\mu\overline{x_{ij}}(t+0.5)+ \frac{1 - \mu}{N_g(N_g -\delta_{IJ})}\sum_{(p,q) \in I \times J, p \neq q}\overline{x_{pq}}(t+0.5)\right)\left(\mu\overline{x_{ji}}(t+0.5)+ \frac{1 - \mu}{N_g(N_g - \delta_{IJ})}\sum_{(r,s) \in I \times J}\overline{x_{rs}}(t+0.5)\right)
\end{dmath}

\begin{dmath}
   \overline{x_{ij}(t+1)x_{ji}(t+1)}   =  \mu^2\overline{x_{ij}(t+0.5)x_{ji}(t+0.5)}+ \frac{\mu(1 - \mu)}{N_g(N_g -1)}\sum_{(p,q) \in I \times J, p \neq q}\overline{x_{pq}(t+0.5)x_{ji}(t+0.5)}  + \frac{\mu(1 - \mu)}{N_g(N_g -\delta_{IJ})}\sum_{(r,s) \in I \times J, r \neq s}\overline{x_{ij}(t+0.5)x_{rs}(t+0.5)}+  \frac{(1 - \mu)^2}{N^2_g(N_g - \delta_{IJ})^2}\sum_{(p,r,q,s) \in I^2 \times J^2, p \neq q, r \neq s}\overline{x_{pq}(t+0.5)x_{rs}(t+0.5)}.
\end{dmath}

For $I = J$, the only case this product is not null, $ \overline{x^2_{IJJI}(t+1)} =  \overline{x^2_{II^jI^jI}(t+1)}$ and:

\begin{dmath}
    \overline{x^2_{II^jI^jI}}(t+1)= \left(\mu^2 + \frac{1 - \mu^2}{N_g(N_g - 1)}\right)\overline{x^2_{II^jI^jI}}(t+0.5) + \left(\frac{(1 - \mu^2)(N_g-2)}{N_g(N_g - 1)}\right)\overline{x^2_{II^jII^p}}(t+0.5) + \left(\frac{(1 - \mu^2)(N_g-2)}{N_g(N_g - 1)}\right)\overline{x^2_{II^jI^pI^j}}(t+0.5) + \left(\frac{(1 - \mu^2)(N_g-2)(N_g-3)}{N_g(N_g - 1)}\right)\overline{x^2_{II^jI^pI^q}}(t+0.5)
    +\left(\frac{1 - \mu^2}{N_g(N_g - 1)}\right)\overline{x^2_{II^jII^j}}(t+0.5) + \left(\frac{2(1 - \mu^2)(N_g-2)}{N_g(N_g - 1)}\right)\overline{x^2_{II^jI^jI^p}}(t+0.5)
\end{dmath}

\subsubsection*{Expression of $x^2_{JIJI}(t+0.5):$}

For $(i,j) \in I \times J, i \neq j$:
\begin{dmath}
    \overline{x_{ji}(t+0.5)x_{ji}(t+0.5)}  =  \overline{x_{ji
    }(t)x_{ji}(t)}  +  \frac{2}{N_c}\left(2\widehat{h_{JI}}(t)(\overline{x_{ii}(t)x_{ji}(t)} - \overline{x_{ji}(t)x_{ji}(t)})  + \widehat{h_{JI}}^2(t)( \overline{x_{ji}(t)x_{ji}(t)} +  \overline{x_{ii}(t)x_{ii}(t)} -  2\overline{x_{ii}(t)x_{ji}(t)}) + \frac{\delta^2}{3} \right) +\frac{2k}{N_T}\sum_{P \in G} \sum_{p \in P, p \notin \{i,j\}}\left(  \widehat{h_{JP}}^2(t)(\overline{x_{pi}(t)x_{pi}(t)} + \overline{x_{ji}(t)x_{ji}(t)}
    -2\overline{x_{pi}(t)x_{ji}(t)} +  \frac{\delta^2}{3}) + 2\widehat{h_{JP}}(t)(\overline{x_{pi}(t)x_{ji}(t)}  - \overline{x_{ji}(t)x_{ji}(t)}) \right).
\end{dmath}

Summing up on all $(i,j) \in I \times J, i \neq j$ and dividing by $N_g (N_g - \delta_{IJ})$, we get:

\begin{dmath}
    \overline{x^2_{JIJI}(t+0.5)}  =  \overline{x^2_{JIJI}(t)} +   \frac{2}{N_c} \left(2\widehat{h_{JI}}(t)( \overline{x^2_{IIJI}(t)} - \overline{x^2_{JIJI}(t)})  + \widehat{h_{JI}}^2(t)( \overline{x^2_{JIJI}(t)} +  \overline{x^2_{IIII}(t)} - 2 \overline{x^2_{IIJI}(t)} + \frac{\delta^2}{3}) \right) + \frac{2k}{N_T}\sum_{P \in G} (N_g - \delta_{IP} - \delta_{JP})\left( \widehat{h_{IP}}^2(t) (\overline{x^2_{PIPI}(t)} + \overline{x^2_{JIJI}(t)})- 2\overline{x^2_{JIPI}(t)} + \frac{\delta^2}{3}) + 2\widehat{h_{JP}}(t)(\overline{x^2_{PIJI}(t)}  - \overline{x^2_{JIJI}(t)}) \right).
\end{dmath}

\subsubsection*{Expression of $x^2_{JIJI}(t+1)$:}

We have:
\begin{dmath}
    \overline{x_{ji}^2}(t+1)=\left( \mu\overline{x_{ji}}(t+0.5) + \frac{1-\mu}{N_g(N_g - \delta_{IJ})} \sum_{(q,p) \in J \times I, q \neq p} \overline{x_{qp}}(t+0.5)\right)^2
\end{dmath}

Hence:
\begin{dmath}
    \overline{x_{ji}^2}(t+1)= \mu^2\overline{x^2_{ji}}(t+0.5) + \frac{2\mu(1-\mu)}{N_g(N_g - \delta_{IJ})} \sum_{(p,q) \in J \times I, p \neq q} \overline{x_{ji}(t+0.5)x_{pq}(t+0.5)} + \frac{(1-\mu)^2}{N^2_g(N_g - \delta_{IJ})^2}\sum_{(p, r, q, s) \in J^2 \times I^2, q \neq p, r \neq s} \overline{x_{pq}(t+0.5)x_{rs}(t+0.5)}.
\end{dmath}

Summing up for all $(i,j) \in I \times J, i \neq j$, and dividing by $N_g(N_g - \delta_{IJ})$ we get:

If $I \neq J$:
\begin{dmath}
    \overline{x^2_{JIJI}}(t+1)= \left(\mu^2 + \frac{1 - \mu^2}{N_g^2}\right)\overline{x^2_{JIJI}}(t+0.5) +  \frac{1-\mu^2}{N_g-1}\overline{x_{JIJ^pI}}(t+0.5).
\end{dmath}

with:
\begin{dmath}
    \overline{x^2_{JIJ^pI^q}}(t) = \frac{1}{N^2_g(N_g - 1)^2} \sum_{(i,p,j, q) \in I^2 \times J^2, i \neq p, j \neq q} \overline{x_{ji}(t)x_{qp}(t)};
\end{dmath}

If $I = J$:

\begin{dmath}
    \overline{x^2_{JIJI}}(t+1)= \left(\mu^2 + \frac{1 - \mu^2}{N_g(N_g - \delta_{IJ})}\right)\overline{x^2_{JIJI}}(t+0.5) + \left(\frac{(1 - \mu^2)(N_g-1-\delta_{IJ})}{N_g(N_g - \delta_{IJ})}\right)\overline{x^2_{JIJI^p}}(t+0.5) + \left(\frac{(1 - \mu^2)(N_g-1-\delta_{IJ})}{N_g(N_g - \delta_{IJ})}\right)\overline{x^2_{JIJ^pI}}(t+0.5) + \left(\frac{(1 - \mu^2)(N_g-1-\delta_{IJ})(N_g-1-2\delta_{IJ})}{N_g(N_g - \delta_{IJ})}\right)\overline{x^2_{JIJ^pI^q}}(t+0.5)
    + \delta_{IJ}\left(
    \left(\frac{1 - \mu^2}{N_g(N_g - \delta_{IJ})}\right)\overline{x^2_{II^jI^jI}}(t+0.5) + \left(\frac{2(1 - \mu^2)(N_g-1-\delta_{IJ})}{N_g(N_g - \delta_{IJ})}\right)\overline{x^2_{II^jI^jI^p}}(t+0.5)
    \right)
\end{dmath}

\subsubsection*{Expression of $x^2_{IJQJ}(t+0.5):$} For $(i,j,q) \in I \times J \times Q, i \neq j, i \neq q, j \neq q$,

\begin{dmath}
    \overline{x_{ij}(t+0.5)x_{qj}(t+0.5)}  =  \overline{x_{ij}(t)x_{qj}(t)}  +  \frac{2}{N_c}\left(\widehat{h_{IJ}}(t)(\overline{x_{jj}(t)x_{qj}(t)} - \overline{x_{ij}(t)x_{qj}(t)}) + \widehat{h_{QJ}}(t)(\overline{x_{jj}(t)x_{ij}(t)} - \overline{x_{ij}(t)x_{qj}(t)} ) \right) +\frac{2k}{N_T}\left( \widehat{h_{IQ}}(t)(\overline{x_{qj}(t)x_{qj}(t)} - \overline{x_{ij}(t)x_{qj}(t)}) + \widehat{h_{QI}}(t)(\overline{x_{ij}(t)x_{ij}(t)} - \overline{x_{ij}(t)x_{qj}(t)}) + \widehat{h_{IQ}}(t)\widehat{h_{QI}}(t)(2\overline{x_{ij}(t)x_{qj}(t)} - \overline{x_{ij}(t)x_{ij}(t)}- \overline{x_{qj}(t)x_{qj}(t)})
    + \sum_{P \in G} \left(\sum_{p \in P, p \notin \{i,j,q\}} \widehat{h_{IP}}(t) (\overline{x_{pj}(t)x_{qj}(t)}
    -\overline{x_{ij}(t)x_{qj}(t)}) + \widehat{h_{QP}}(t)(\overline{x_{pj}(t)x_{ij}(t)}
    -\overline{x_{ij}(t)x_{qj}(t)}) \right)\right).
\end{dmath}

Summing up on all $(i,j,q) \in I \times J \times Q, i \neq j, i \neq q, j \neq q$ and dividing by $N_g (N_g - \delta_{IJ})(N_g - \delta_{IQ} - \delta_{JQ})$, we get:

\begin{dmath}
    \overline{x^2_{IJQJ}(t+0.5)}  =  \overline{x^2_{IJQJ}(t)} +   \frac{2}{N_c} \left(\widehat{h_{IJ}}(t)( \overline{x^2_{JJQJ}(t)} - \overline{x^2_{IJQJ}(t)}) + \widehat{h_{QJ}}(t)( \overline{x^2_{JJIJ}(t)} - \overline{x^2_{IJQJ}(t)})  \right) 
    +\frac{2k}{N_T}\left( \widehat{h_{IQ}}(t)(\overline{x^2_{QJQJ}(t)} - \overline{x^2_{IJQJ}(t)} ) + \widehat{h_{QI}}(t)(\overline{x^2_{IJIJ}(t)} - \overline{x^2_{IJQJ}(t)}) + \widehat{h_{IQ}}(t)\widehat{h_{QI}}(t)(2\overline{x^2_{IJQJ}(t)} - \overline{x^2_{IJIJ}(t)}- \overline{x^2_{QJQJ}(t)})
    + \sum_{P \in G} (Ng - \delta_{PI} - \delta_{PJ} - \delta_{PQ}) \left( \widehat{h_{IP}}(t) (\overline{x^2_{PJQJ}(t)}
    -\overline{x^2_{IJQJ}(t)}) + \widehat{h_{QP}}(t)(\overline{x^2_{PJIJ}(t)}
    -\overline{x^2_{IJQJ}(t)})\right)\right).
\end{dmath}

\subsubsection*{Expression of $x^2_{IJQJ}(t+1)$:}

We assume $ J \neq I, Q \neq I$. For $(i,j,q) \in I \times J \times Q$:
\begin{dmath}
   \overline{x_{ji}(t+1)x_{qi}(t+1)}  =  \mu^2\overline{x_{ji}(t+0.5)x_{qi}(t+0.5)}+ \frac{\mu (1 - \mu)}{N_g(N_g-\delta_{IJ})}\sum_{(p,r) \in J \times I}\overline{x_{pr}(t+0.5)x_{qi}(t+0.5)} + \frac{\mu (1 - \mu)}{N_g(N_g-\delta_{IQ})}\sum_{(s,t) \in Q \times I} \overline{x_{ji}(t+0.5)x_{st}(t+0.5)} +  \frac{(1 - \mu)^2}{N^2_g(N_g-\delta_{IJ})(N_g-\delta_{IQ})}\sum_{(p,s,r,t) \in J \times Q \times I^2}\overline{x_{pr}(t+0.5)x_{st}(t+0.5)}.
\end{dmath}

Summing up over all $(i,j,q) \in I \times J \times Q$ and dividing by $N_g^2(N_g-\delta_{IJ})(N_g-\delta_{IQ})$, we get:

\begin{dmath}
    \overline{x^2_{JIQI}}(t+1)= \left(\mu^2 + \frac{(1 - \mu^2)(N_g-\delta_{IQ}-\delta_{JQ})}{N_g(N_g - \delta_{IQ})}\right)\overline{x^2_{JIQI}}(t+0.5) 
    + \left(\frac{(1 - \mu^2)(N_g-\delta_{IQ}-\delta_{JQ})(N_g-\delta_{IJ}-\delta_{IQ}-1)}{N_g(N_g-\delta_{IQ})}\right)\overline{x^2_{JIQI^p}}(t+0.5) 
    +\delta_{IQ}\left(\frac{(1 - \mu^2)(N_g-\delta_{IJ}-1)}{N_g(N_g - \delta_{IQ})}\right)\overline{x^2_{JIII^p}}(t+0.5) 
    +\delta_{IJ}\left(\frac{(1 - \mu^2)(N_g-\delta_{IQ}-1)}{N_g(N_g - \delta_{IJ})}\right)\overline{x^2_{II^pQI}}(t+0.5) 
    +\delta_{JQ}\left(\frac{(1 - \mu^2)}{N_g(N_g - \delta_{IQ})}\right)\overline{x^2_{JIJI}}(t+0.5) 
    +\delta_{JQ}\left(\frac{(1 - \mu^2)(N_g - \delta_{IJ} -1)}{N_g(N_g - \delta_{IQ})}\right)\overline{x^2_{JIJI^p}}(t+0.5) 
    +\delta_{IJ}\delta_{IQ}\left(\frac{(1 - \mu^2)}{N_g(N_g - 1)}\right)\overline{x^2_{II^jI^jI}}(t+0.5) 
\end{dmath}

\subsubsection*{Expression of $x^2_{IJQI}(t+0.5):$}
This product is null for $I \neq J$.
For $(i,j, q) \in I \times J \times Q, i \neq j, i \neq q, j \neq q$:
\begin{dmath}
    \overline{x_{ij}(t+0.5)x_{qi}(t+0.5)}  =  \overline{x_{ij}(t)x_{qi}(t)}  +  \frac{2}{N_c}\left(\widehat{h_{IJ}}(t)(\overline{x_{jj}(t)x_{qi}(t)} - \overline{x_{ij}(t)x_{qi}(t)}) + \widehat{h_{QI}}(t)(\overline{x_{ii}(t)x_{ij}(t)} - \overline{x_{ij}(t)x_{qi}(t)} )\right) +\frac{2k}{N_T}\sum_{P \in G} \left(\sum_{p \in P, p \notin \{i,q\}} \widehat{h_{QP}}(t) (\overline{x_{ij}(t)x_{pi}(t)}
    -\overline{x_{ij}(t)x_{qi}(t)}) +\sum_{p \in P, p \notin \{i,j\}} \widehat{h_{IP}}(t)(\overline{x_{qi}(t)x_{pj}(t)}
    -\overline{x_{qi}(t)x_{ij}(t)}) \right).
\end{dmath}

Summing up on all $(i,j,q) \in I \times J \times Q, i \neq j, i \neq q, j \neq q$ and dividing by $N_g (N_g - \delta_{IJ})(N_g - \delta_{IQ} - \delta_{JQ})$, we get:

\begin{dmath}
    \overline{x^2_{IJQI}(t+0.5)}  =  \overline{x^2_{IJQI}(t)} +   \frac{2}{N_c} \left(\widehat{h_{IJ}}(t)( \overline{x^2_{JJQI}(t)} - \overline{x^2_{IJQI}(t)}) + \widehat{h_{QI}}(t)( \overline{x^2_{IIIJ}(t)} - \overline{x^2_{IJQI}(t)})  \right)
    + \frac{2k}{N_T}\sum_{P \in G}\left( (N_g - \delta_{IP} - \delta_{QP})\widehat{h_{QP}}(t) (\overline{x^2_{IJPI}(t)} - \overline{x^2_{IJQI}(t)}) +   (N_g - \delta_{IP} - \delta_{JP})\widehat{h_{IP}}(t) (\overline{x^2_{QIPJ}(t)} - \overline{x^2_{IJQI}(t)}) \right).
\end{dmath}

\subsubsection*{Expression of $x^2_{IJQI}(t+1)$:}

\begin{dmath}
    \overline{x^2_{JIII^p}}(t+1)= \left(\mu^2 + 
    \frac{(1 - \mu^2)(N_g-\delta_{IJ}-1)}{N_g(N_g - 1)} \right)\overline{x^2_{JIII^p}}(t+0.5) 
    + \left(\frac{(1 - \mu^2)(N_g-1-\delta_{IJ})}{N_g(N_g - 1)}\right)\overline{x^2_{JII^pI}}(t+0.5) 
    +\left(\frac{(1 - \mu^2)(N_g-1-\delta_{IJ})(N_g-2-\delta_{IJ})}{N_g(N_g-1)} \right)\overline{x^2_{JII^pI^q}}(t+0.5) 
    +\delta_{IJ}\left(\frac{(1 - \mu^2)(N_g-2)}{N_g(N_g - 1)}\right)\overline{x^2_{II^pI^qI}}(t+0.5) 
    +\delta_{IJ}\left(\frac{(1 - \mu^2)}{N_g(N_g - 1)}\right)\overline{x^2_{II^pII^p}}(t+0.5) 
    +\delta_{IJ}\left(\frac{(1 - \mu^2)(N_g - 2)}{N_g(N_g - 1)}\right)\overline{x^2_{II^pII^q}}(t+0.5) 
    +\delta_{IJ}\left(\frac{(1 - \mu^2)}{N_g(N_g - 1)}\right)\overline{x^2_{II^jI^jI}}(t+0.5) 
\end{dmath}

\subsubsection*{Expression of $x^2_{IJIQ}(t+0.5):$}
This product is null for $J \neq Q$.
For $(i,j, q) \in I \times J \times Q, i \neq j, i \neq q, j \neq q$:
\begin{dmath}
    \overline{x_{ij}(t+0.5)x_{iq}(t+0.5)}  =  \overline{x_{ij}(t)x_{iq}(t)}  +  \frac{2}{N_c}\left(\widehat{h_{IJ}}(t)(\overline{x_{jj}(t)x_{iq}(t)} - \overline{x_{ij}(t)x_{iq}(t)}) + \widehat{h_{IQ}}(t)(\overline{x_{qq}(t)x_{ij}(t)} - \overline{x_{ij}(t)x_{iq}(t)} )\right) +\frac{2k}{N_T}\sum_{P \in G} \left(\sum_{p \in P, p \notin \{i,q\}} \widehat{h_{IP}}(t) (\overline{x_{ij}(t)x_{pq}(t)}
    -\overline{x_{ij}(t)x_{iq}(t)}) +\sum_{p \in P, p \notin \{i,j\}}\widehat{h_{IP}}(t)(\overline{x_{iq}(t)x_{pj}(t)}
    -\overline{x_{iq}(t)x_{ij}(t)}) \right).
\end{dmath}

Summing up on all $(i,j, q) \in I \times J \times Q, i \neq j, i \neq q, j \neq q$ and dividing by $N_g (N_g - \delta_{IJ})(N_g - \delta_{IQ} - \delta_{JQ})$, we get:

\begin{dmath}
    \overline{x^2_{IJIQ}(t+0.5)}  =  \overline{x^2_{IJIQ}(t)} +   \frac{2}{N_c} \left(\widehat{h_{IJ}}(t)( \overline{x^2_{JJIQ}(t)} - \overline{x^2_{IJIQ}(t)}) + \widehat{h_{IQ}}(t)( \overline{x^2_{QQIJ}(t)} - \overline{x^2_{IJIQ}(t)})\right) + \frac{2k}{N_T}\left(\sum_{P \in G} (N_g - \delta_{IP} - \delta_{QP}) \widehat{h_{IP}}(t) (\overline{x^2_{IJPQ}(t)} - \overline{x^2_{IJIQ}(t)}) +   (N_g - \delta_{IP} - \delta_{JP}) \widehat{h_{IP}}(t) (\overline{x^2_{PJIQ}(t)} - \overline{x^2_{IJIQ}(t)}) \right).
\end{dmath}

\subsubsection*{Expression of $x^2_{IJIQ}(t+1)$:}

For $(j,i,q) \in J \times I \times Q$:

\begin{dmath}
   \overline{x_{ji}(t+1)x_{jq}(t+1)}   =  \left(\mu\overline{x_{ji}}(t+0.5)+ \frac{1 - \mu}{N_g(N_g-\delta_{IJ})}\sum_{(r,s) \in I \times J, r \neq s}\overline{x_{rq}}(t+0.5)\right)\left(\mu\overline{x_{jq}}(t+0.5)+ \frac{1 - \mu}{N_g(N_g - \delta_{IQ})}\sum_{(m,n) \in I \times Q, m \neq n}\overline{x_{mn}}(t+0.5)\right)
\end{dmath}

\begin{dmath}
   \overline{x_{ji}(t+1)x_{jq}(t+1)}   =  \mu^2\overline{x_{ij}(t+0.5)x_{ji}(t+0.5)}+ \frac{\mu(1 - \mu)}{N_g(N_g - \delta_{IJ})}\sum_{(r,s) \in I \times J, r \neq s }\overline{x_{rs}(t+0.5)x_{jq}(t+0.5)}  + \frac{\mu(1 - \mu)}{N_g(N_g -\delta_{IQ})}\sum_{(m,n) \in I \times Q, m \neq n}\overline{x_{ji}(t+0.5)x_{mn}(t+0.5)}+  \frac{(1 - \mu)^2}{N^2_g(N_g - \delta_{IJ})(N_g - \delta_{IQ})}\sum_{(r,s,m,n) \in I \times J \times I \times Q, r \neq s, m \neq n}\overline{x_{rs}(t+0.5)x_{mn}(t+0.5)}.
\end{dmath}

For $I = Q$, the only case this product is not null, $ \overline{x^2_{JIJQ}(t+1)} =  \overline{x^2_{JIJI^p}(t+1)}$. Summing up over all $(j,i,q) \in J \times I^2, j \neq i, j \neq q$ and dividing by $N_g(N_g - \delta_{IJ})^2$, we get:

\begin{dmath}
    \overline{x^2_{JIJI^p}}(t+1)= \left(\mu^2 + \frac{(1 - \mu^2)(N_g-1-\delta_{IJ})}{N_g(N_g - \delta_{IJ})}\right)\overline{x^2_{JIJI^p}}(t+0.5) + \left(\frac{1 - \mu^2}{N_g(N_g - \delta_{IJ})}\right)\overline{x^2_{JIJI}}(t+0.5) + \left(\frac{(1 - \mu^2)(N_g-1-\delta_{IJ})}{N_g(N_g - \delta_{IJ})}\right)\overline{x^2_{JIJ^pI}}(t+0.5) + \left(\frac{(1 - \mu^2)(N_g-1-\delta_{IJ})(N_g-1-2\delta_{IJ})}{N_g(N_g - \delta_{IJ})}\right)\overline{x^2_{JIJ^pI^q}}(t+0.5)
    + \delta_{IJ}\left(
    \left(\frac{1 - \mu^2}{N_g(N_g - \delta_{IJ})}\right)\overline{x^2_{II^jI^jI}}(t+0.5) + \left(\frac{2(1 - \mu^2)(N_g-1-\delta_{IJ})}{N_g(N_g - \delta_{IJ})}\right)\overline{x^2_{II^jI^jI^p}}(t+0.5)
    \right)
\end{dmath}

\subsubsection*{Expression of $x^2_{IIJQ}(t+0.5):$}
This product is null for $I \neq Q$.
For $(i,j, q) \in I \times J \times Q, i \neq j, i \neq q, j \neq q$:

\begin{dmath}
    \overline{x^2_{ii}}(t+0.5)  =  \overline{x^2_{ii}}(t) + \frac{2}{N_c} \sum_{J \in G} \sum_{j \in J, j \neq i} \left((\widehat{h^2_{IJ}}(t)-2\widehat{h_{IJ}}(t))\overline{x^2_{ii}}(t) + \widehat{h_{IJ}}(t)^2\overline{x^2_{ji}}(t) + 2 (1-\widehat{h_{IJ}}(t))\widehat{h_{IJ}}(t)\overline{x_{ii}(t)x_{ji}(t)} + \widehat{h_{IJ}}(t)^2 \frac{\delta^2}{3} \right).
\end{dmath}

Summing up on all $(i,jbq) \in I \times J \times Q, i \neq j, i \neq q, j \neq q$ and dividing by $N_g (N_g - \delta_{IJ})(N_g - \delta_{IQ} - \delta_{JQ})$, we get:

\begin{dmath}
    \overline{x^2_{IIII}}(t+0.5)  = \overline{x^2_{IIII}}(t) + \sum_{J \in G}\frac{2(N_g - \delta_{IJ})}{N_c} \left((\widehat{h^2_{IJ}}(t)-2\widehat{h_{IJ}}(t))\overline{x^2_{IIII}}(t) + \widehat{h_{IJ}}(t)^2\overline{x^2_{JIJI}}(t) + 2 (1-\widehat{h_{IJ}}(t))\widehat{h_{IJ}}(t)\overline{x^2_{IIJI}(t)} + \widehat{h_{IJ}}(t)^2 \frac{\delta^2}{3} \right).
\end{dmath}

\subsubsection*{Expression of $x^2_{IIJQ}(t+1)$:}

For $(i,j,q) \in I \times J \times Q$:

\begin{dmath}
   \overline{x_{ii}(t+1)x_{jq}(t+1)}   =  \left(\mu\overline{x_{ii}}(t+0.5)+ \frac{1 - \mu}{N_g}\sum_{(p) \in I}\overline{x_{pp}}(t+0.5)\right)\left(\mu\overline{x_{jq}}(t+0.5)+ \frac{1 - \mu}{N_g(N_g - \delta_{JQ})}\sum_{(r,s) \in J \times Q, r \neq s}\overline{x_{rs}}(t+0.5)\right)
\end{dmath}

\begin{dmath}
   \overline{x_{ii}(t+1)x_{jq}(t+1)}   =  \mu^2\overline{x_{ii}(t+0.5)x_{jq}(t+0.5)}+ \frac{\mu(1 - \mu)}{N_g}\sum_{p \in I }\overline{x_{pp}(t+0.5)x_{jq}(t+0.5)}  + \frac{\mu(1 - \mu)}{N_g(N_g -\delta_{JQ})}\sum_{(r,s) \in I \times J, r \neq s}\overline{x_{ii}(t+0.5)x_{rs}(t+0.5)}+  \frac{(1 - \mu)^2}{N^2_g(N_g - \delta_{IJ})}\sum_{(p,r,s) \in I \times J \times Q, r \neq s}\overline{x_{pp}(t+0.5)x_{rs}(t+0.5)}.
\end{dmath}

For $Q = I$, the only case this product is not null, $ \overline{x^2_{IIJQ}(t+1)} =  \overline{x^2_{IIJI^p}(t+1)}$. Summing up over all $(i,q,j) \in I^2 \times J, q \neq j$ and dividing by $N_g^2(N_g - \delta_{IJ})$, we get approximately the same formula that $\overline{x^2_{IIJI}(t+1)}$:

\begin{dmath}
   \overline{x^2_{IIJI^p}(t+1)}   = \left( \mu^2+ \frac{(1 - \mu^2)(N_g-1-\delta_{IJ})}{N_g}\right)\overline{x^2_{IIJI^p}(t+0.5)} + \frac{(1 - \mu^2)}{N_g} \overline{x^2_{IIJI}(t+0.5)} + \frac{\delta_{IJ}(1 - \mu^2)}{N_g} \overline{x^2_{IIII^q}(t+0.5)}.
\end{dmath}

\subsubsection*{Expression of $x^2_{IJQR}(t+0.5):$}
This product is null for $J \neq R$.
For $(i,j, q, r) \in I \times J \times Q \times R, i \neq j, i \neq q, i \neq r, j \neq q, j \neq r, q \neq r$:
\begin{dmath}
    \overline{x_{ij}(t+0.5)x_{qr}(t+0.5)}  =  \overline{x_{ij}(t)x_{qr}(t)}  +  \frac{2}{N_c}\left(\widehat{h_{IJ}}(t)(\overline{x_{jj}(t)x_{qr}(t)} - \overline{x_{ij}(t)x_{qr}(t)}) + \widehat{h_{QR}}(t)(\overline{x_{rr}(t)x_{ij}(t)} - \overline{x_{ij}(t)x_{qr}(t)} ) \right) +\frac{2k}{N_T}\sum_{P \in G} \left(\sum_{p \in P, p \notin \{q,r\}} \widehat{h_{QP}}(t) (\overline{x_{ij}(t)x_{pr}(t)}
    -\overline{x_{ij}(t)x_{qr}(t)}) +\sum_{p \in P, p \notin \{i,j\}}\widehat{h_{IP}}(t)(\overline{x_{qr}(t)x_{pj}(t)}
    -\overline{x_{qr}(t)x_{ij}(t)}) \right).
\end{dmath}

Summing up on all $(i,j, q, r) \in I \times J \times Q, i \neq j, i \neq q, i \neq r, j \neq q, j \neq r, q \neq r$ and dividing by $N_g (N_g - \delta_{IJ})(N_g - \delta_{IQ} - \delta_{JQ})(N_g - \delta_{IR} - \delta_{JR} - \delta_{QR})$, we get:

\begin{dmath}
    \overline{x^2_{IJQR}(t+0.5)}  =  \overline{x^2_{IJQR}(t)} +   \frac{2}{N_c} \left(\widehat{h_{IJ}}(t)( \overline{x^2_{JJQR}(t)} - \overline{x^2_{IJQR}(t)}) + \widehat{h_{QR}}(t)( \overline{x^2_{RRIJ}(t)} - \overline{x^2_{IJQR}(t)}) \right) + \frac{2k}{N_T}\sum_{P \in G}\left( (N_g - \delta_{QP} - \delta_{RP}) \widehat{h_{QP}}(t) (\overline{x^2_{IJPR}(t)} - \overline{x^2_{IJQR}(t)}) +  (N_g - \delta_{IP} - \delta_{JP}) \widehat{h_{IP}}(t) (\overline{x^2_{PJQR}(t)} - \overline{x^2_{IJQR}(t)}) \right).
\end{dmath}

Now, we express how these quantities change because of the process of attraction of the average opinion.

\subsubsection*{Expression of $x^2_{IJQR}(t+1)$:}

We assume $ J \neq I, Q \neq I$. For $(j,i,q, m) \in J \times I \times Q \times I$:
\begin{dmath}
   \overline{x_{ji}(t+1)x_{qm}(t+1)}  =  \mu^2\overline{x_{ji}(t+0.5)x_{qm}(t+0.5)}+ \frac{\mu (1 - \mu)}{N_g(N_g-\delta_{JI})}\sum_{(p,r) \in J \times I, p \neq r}\overline{x_{pr}(t+0.5)x_{qm}(t+0.5)} + \frac{\mu (1 - \mu)}{N_g(N_g-\delta_{QI})}\sum_{(s,t) \in Q \times I, s \neq t} \overline{x_{ji}(t+0.5)x_{st}(t+0.5)} +  \frac{(1 - \mu)^2}{N^2_g(N_g-\delta_{JI})(N_g-\delta_{QI})}\sum_{(p,s,r,t) \in J \times Q \times I^2, p \neq r, s \neq t}\overline{x_{pr}(t+0.5)x_{st}(t+0.5)}.
\end{dmath}

Summing up over all $(j,i,q, m) \in J \times I \times Q \times q, i \neq m$ and dividing by $N_g^3(N_g -1)$, we get:

\begin{dmath}
    \overline{x^2_{JIQI^p}}(t+1)= \left(\mu^2 + \frac{(1 - \mu^2)(N_g-\delta_{IQ}-\delta_{JQ})(N_g-\delta_{IJ}-\delta_{IQ}-1)}{N_g(N_g-\delta_{IQ})}\right)\overline{x^2_{JIQI^p}}(t+0.5) 
    + \left(\frac{(1 - \mu^2)(N_g-\delta_{IQ}-\delta_{JQ})}{N_g(N_g - \delta_{IQ})}\right)\overline{x^2_{JIQI}}(t+0.5) 
    +\delta_{IQ}\left(\frac{(1 - \mu^2)(N_g-\delta_{IJ}-1)}{N_g(N_g - \delta_{IQ})}\right)\overline{x^2_{JIII^p}}(t+0.5) 
    +\delta_{IJ}\left(\frac{(1 - \mu^2)(N_g-\delta_{IQ}-\delta_{JQ})}{N_g(N_g - \delta_{IJ})}\right)\overline{x^2_{II^pQI}}(t+0.5) 
    +\delta_{JQ}\left(\frac{(1 - \mu^2)}{N_g(N_g - \delta_{IQ})}\right)\overline{x^2_{JIJI}}(t+0.5) 
    +\delta_{JQ}\left(\frac{(1 - \mu^2)(N_g - \delta_{IJ} -1)}{N_g(N_g - \delta_{IQ})}\right)\overline{x^2_{JIJI^p}}(t+0.5) 
    +\delta_{IJ}\delta_{IQ}\left(\frac{(1 - \mu^2)}{N_g(N_g - 1)}\right)\overline{x^2_{II^jI^jI}}(t+0.5) 
\end{dmath}

\end{document}